%% file: betha-mainpaper.tex
\documentclass[journal]{IEEEtran}

\usepackage{graphicx,amssymb,amstext,amsmath,color}
\DeclareGraphicsExtensions{.eps,.pdf,.png,.jpg,.gif,.jpeg,.pstex}

\input{epsf.sty}

\usepackage{times, epsf}
\usepackage{amsmath,amssymb}
\usepackage{url}
\usepackage[font = footnotesize]{caption}
\usepackage{ float}
\usepackage{algorithm}
\usepackage{algorithmic}
\usepackage{graphicx}
\usepackage{subfig}
\usepackage{cite}
\usepackage{empheq}
\usepackage{multirow,tabularx}
\usepackage[pdftex]{epsfig}
\input{epsf.sty}

\usepackage{stackrel}
\usepackage{amsfonts}
\input{epsf.sty}

\usepackage{times, epsf}
\usepackage{graphicx}
\usepackage{cite}
\usepackage{multirow,tabularx}
\usepackage{epstopdf}

\newtheorem{thm}{Theorem}
\newtheorem{lemma}{Lemma}
\newtheorem{cor}{Corollary}
\newtheorem{defn}{Definition}
\newtheorem{remark}{\indent \bf Remark}

\newcommand{\expect}[1]{{\mathbb{E}}\left[{#1}\right]}

\newcommand{\E}[1]{\expect{#1}}

\newcommand{\ba}{\begin{array}}
\newcommand{\ea}{\end{array}}

\def\PARstart#1#2{\begingroup\def\par{\endgraf\endgroup\lineskiplimit=0pt}
    \setbox2=\hbox{\uppercase{#2} }\newdimen\tmpht \tmpht \ht2
    \advance\tmpht by \baselineskip\font\hhuge=cmr10 at \tmpht
    \setbox1=\hbox{{\hhuge #1}}
    \count7=\tmpht \count8=\ht1\divide\count8 by 1000 \divide\count7 by\count8
    \tmpht=.001\tmpht\multiply\tmpht by \count7\font\hhuge=cmr10 at \tmpht
    \setbox1=\hbox{{\hhuge #1}} \noindent \hangindent1.05\wd1
    \hangafter=-2 {\hskip-\hangindent \lower1\ht1\hbox{\raise1.0\ht2\copy1}%
    \kern-0\wd1}\copy2\lineskiplimit=-1000pt}

\newcommand{\bit}{\begin{itemize}}
\newcommand{\eit}{\end{itemize}}

\input{bethamacros}

\begin{document}
\title{{Adaptive Video Streaming for Wireless Networks with Multiple Users and Helpers}}
\author{\IEEEauthorblockN{Dilip Bethanabhotla, \IEEEmembership{Student Member,~IEEE,} Giuseppe Caire, \IEEEmembership{Fellow,~IEEE,} \\ and Michael J. Neely, \IEEEmembership{Senior Member,~IEEE}}
\thanks{This research has been partially supported by Intel, Cisco and Verizon Wireless in the framework of the VAWN (Video Aware Wireless Networks) project. The authors are with the Department of Electrical Engineering,  University of Southern California, Los Angeles CA. Email: {\tt bethanab, caire, mjneely@usc.edu}}}

\maketitle

\vspace{-2.2cm}

\begin{abstract}
We consider the design of a scheduling policy for video streaming 
in a wireless network formed by several users and helpers (e.g., base stations).
In such networks, any user is typically in the range of multiple helpers. 
Hence, an efficient policy should allow the users to dynamically select the helper nodes 
to download from and determine adaptively the quality level of the requested video segment.
In order to obtain a tractable formulation, we follow a ``divide and conquer'' approach: 
i) We formulate a Network Utility Maximization (NUM) problem where
the network utility function is a concave and componentwise non-decreasing function of the time-averaged 
users' requested video quality index and maximization is subject to the stability of all queues in the system. 
ii) We solve the NUM problem by using a Lyapunov Drift Plus Penalty approach, 
obtaining a dynamic adaptive scheme that decomposes into two building blocks: 
1) adaptive video quality and helper selection (run at the user nodes); 
2) dynamic allocation of the helper-to-user transmission rates (run at the help nodes). 
Our solution provably achieves NUM optimality in a strong per-sample path sense (i.e., without assumptions of stationarity and ergodicity). 
iii) We observe that, since all queues in the system are stable, all requested video chunks shall be eventually delivered.
iv) In order to translate the requested video quality into the effective video quality at the user playback, it is necessary that
the chunks are delivered within their playback deadline. This requires that the largest delay among all queues at the helpers 
serving any given user is less than the pre-buffering time of that user at its streaming session startup phase. 
In order to achieve this condition with high probability, we propose an effective and decentralized (albeit heuristic) scheme 
to adaptively calculate the pre-buffering and re-buffering time at each user. 
In this way, the system is forced to work in the ``smooth streaming regime,'' i.e., in the regime of very small playback buffer underrun rate.
Through simulations, we evaluate the performance of the proposed algorithm under realistic assumptions of a network with densely 
deployed helper and user nodes, including  user mobility, variable bit-rate video coding, and users joining or leaving the system  
at arbitrary times.
\end{abstract}




\section{Introduction}  \label{sec:intro-dilip}

Wireless data traffic is predicted to increase dramatically in the next few years, 
up to two orders of magnitude by 2020 \cite{cisco66}.
This increase is mainly due to streaming of Video on Demand (VoD), enabled by 
multimedia devices such as tablets and smartphones.
It is well understood that the current trend of cellular technology (e.g., LTE \cite{sesia-LTE}) 
cannot cope with such traffic increase, unless the density of the deployed wireless infrastructure is increased correspondingly. 
This motivates the recent flurry of research on massive and dense deployment of base station antennas, either in the form of
``massive MIMO'' solutions (hundreds of antennas at each cell site \cite{Marzetta-TWC10,Huh11,hoydis2011massive})
or in the form of very dense small-cell networks (multiple nested tiers of 
smaller and smaller cells, possibly operating at higher and higher carrier frequencies \cite{chandrasekhar2008femtocell, hoydis2011green}). 
While discussing the relative merits of these approaches is out of the scope of this paper, we mention here that the 
small-cell solution appears to be particularly attractive to handle a high density of nomadic (low mobility) users 
demanding high data rates,  as for typical VoD streaming users.

Motivated by these considerations, in this paper we envisage a network 
formed by densely deployed fixed nodes (hereafter denoted as {\em helpers}), serving multiple stationary or low-mobility (nomadic) 
video-streaming users. We focus on VoD streaming, where users start their streaming sessions at random times, 
and demand different video files. Hence, the approach of having all users overhearing a common multicasting data stream, as in live streaming, 
is not applicable. In contrast, each streaming user requests sequentially a number of video segments (referred to as {\em chunks})
and starts its playback after some {\em pre-buferring} delay, typically much smaller than the duration of the whole streaming session. 
In order to guarantee continuous playback in the streaming session, the system has to ensure that each video chunk
is delivered before its playback deadline. This fundamentally differentiates {\em VoD streaming} from both 
{\em live streaming} and {\em file downloading}.

This paper focuses on the design of a scheduling policy for VoD streaming
in a wireless network formed by many users and helpers, deployed over a localized geographic area and sharing the same channel bandwidth.  
We focus on the wireless segment of the network, assuming that the video files
are already present at the helper nodes. This condition holds when the backhaul connecting the helper nodes to some 
video server in the core network is fast enough, such that we  can neglect the delays introduced by the backhaul. 
In the case where such fast backhaul is not present, the recently proposed approach of caching 
at the wireless edge (see \cite{mingyue-isit13,mingyue-JSAC,ComMag,femtocaching-INFOCOM,femtocaching-ICC,
maddah2013fundamental,maddah2013decentralized,niesen2013coded,pedarsani2013online,
ji2014order,ji2013fundamental}) was shown to be able to exploit the inherent asynchronous content reuse of VoD 
in order to predict and proactively store the popular video files such that, with high probability, the demanded files 
are effectively already present in the helpersÕ caches. This justifies our assumption of neglecting the effects of the wired backhaul 
and focusing only on the wireless segment of the system.

{\bf Contributions:}  
In order to obtain a tractable formulation,  we follow a ``divide and conquer'' approach, 
conceptually organized in the following steps: 

i) We formulate a Network Utility Maximization (NUM) problem \cite{kelly2006mathematics,yi2008stochastic,chiang2007layering}
where the network utility function is a concave and componentwise non-decreasing function of the time-averaged 
users' {\em requested} video quality index and the maximization is subject to the stability of all queues in the system. The shape of the network utility function
can be chosen in order to enforce some desired notion of fairness across the users \cite{mo2000fair}.

ii) We solve the NUM problem in the framework of Lyapunov Optimization \cite{neely2010stochastic}, using the {\em drift plus penalty} (DPP) 
approach~\cite{neely2010stochastic}.  
The obtained solution is provably asymptotically optimal (with respect to the defined NUM problem)
on a per-sample path sense  (i.e., without assuming stationarity and ergodicity of the underlying network state process \cite{neely2010stochastic,neely2010universal}).
Furthermore, it naturally decomposes into sub-policies that can be implemented in a distributed way, 
by functions performed at the users and the helpers, requiring only {\em local} information.  
The function implemented at the user nodes is referred to as {\em congestion control,} since it consists of the 
adaptive selection of the video quality and the serving helper.  
The function implemented at the helpers is referred to as {\em transmission scheduling,} since it corresponds to the adaptive 
selection of  the user to be served on the downlink of each helper station. 

iii) We observe that, since all queues in the system are stable, all {\em requested} video chunks shall be eventually {\em delivered}.

iv) As a consequence, in order to ensure that all the video chunks are delivered within their playback deadline, it is sufficient to ensure that 
the largest delay among all queues at the helpers serving any given user is not larger than the pre-buffering time allowed for that user
at its streaming session startup phase. We refer to the event that a chunk is not delivered within its playback deadline as a 
{\em buffer underrun} event. Since such events are perceived as very harmful for the overall quality of the streaming session, 
the system must operate in the regime where the relative fraction of such chunks (referred to as {\em buffer underrun rate}) is small. 
We refer to such desirable regime as the {\em smooth streaming regime}. 
 In particular, when the maximum delay of each queue in the system admits a deterministic upper bound (e.g., see \cite{neely2012wireless}), 
setting the pre-buffering time larger than such bound makes the underrun rate equal to zero. 
However, for a system with arbitrary user mobility, 
arbitrary per-chunk fluctuations of the video coding rate (as in typical Variable Bit-Rate (VBR) coding \cite{ortega2000variable}), 
and users joining or leaving the system at arbitrary times, such deterministic delay upper bounds do not exist. 
Hence, in order to make the system operate in the smooth streaming regime, we propose a method to locally estimate the delays with which 
the video packets  are delivered, such that each user can calculate its pre-buffering and re-buffering time to be larger than the locally estimated
maximum queue delay. Through simulations, we demonstrate that the combination of our scheduling policy and adaptive pre-buffering scheme 
is able to achieve the desired fairness across the users and, at the same time, very small playback buffer underrun rate. 

Since the proposed policy achieves NUM optimality on a per-sample path basis and,
thanks to the adaptive dimensioning of the users' pre-buffering time, the system operates in the regime of small buffer underrun rate
(i.e., in the smooth streaming regime),  
the resulting system performance is near-optimal in the following sense: for any {\em bounded} penalty weight 
assigned to the buffer underrun events, the system network utility (including such penalty) is just a small perturbation away from
the optimal NUM value obtained by our DPP policy (see Remark \ref{rem-smooth-streaming}).
 

The rest of this paper is organized as follows.  
In Section \ref{sec:sysmodel-dilip}, we describe the system model for VoD streaming 
in a { wireless network with multiple users and helpers}, 
and discuss some key underlying assumptions.  
In Section \ref{sec:numiidstate-dilip}, we  formulate the NUM problem, provide
the proposed distributed dynamic scheduling policy for its solution, and state the main results on its optimality. 
Section \ref{sec:prebuffering} illustrates our proposed scheme for adaptive pre-buffering and re-buffering in order to cope with playback buffer underrun events.
Finally, simulation results illustrating the particular features of the proposed scheme  are provided in 
Section \ref{sec:simul-dilip}.  The main technical proofs are collected in the Appendices, 
in order to maintain the flow of exposition. 

\section{System Model} \label{sec:sysmodel-dilip}

We consider a discrete, time-slotted wireless network with multiple users and multiple helper stations sharing the same bandwidth. 
The network is defined by a bipartite graph $\Gc = (\Uc, \Hc, \Ec)$, where $\Uc$ denotes the set of users, $\Hc$ denotes the set of helpers, and 
$\Ec$ contains edges for all pairs $(h,u)$ such that there exists a potential transmission 
link between $h \in \Hc$ and $u \in \Uc$.\footnote{The existence of such potential links depends on the channel gain coefficients between 
helper $h$ and user $u$ (see the physical channel model in Section \ref{phy-model}), as well as on some protocol imposing
restricted access for some helpers.}  
We denote by $\Nc(u) \subseteq \Hc$ the neighborhood of user $u$, i.e., 
$\Nc(u) = \{ h \in \Hc : (h,u) \in \Ec\}$. Similarly,  $\Nc(h) = \{u \in \Uc : (h,u) \in \Ec\}$. 

Each user $u \in \Uc$ requests a video file $f_u$ from a library $\Fc$ of possible files. 
Each video file is formed by a sequence of  chunks. 
Each chunk corresponds to a group of pictures (GOP) that are encoded and decoded 
as stand-alone units~\cite{sanchez2011idash}.  Chunks have a fixed playback duration, 
given by $T_{\rm gop} = \mbox{(\# frames per GOP)}/\eta$, where $\eta$ is the frame rate, expressed in  
frames per second.  The streaming process consists of transferring chunks from the helpers to the requesting 
users such that the playback buffer at each user contains the required chunks at the beginning 
of each chunk playback deadline. The playback starts after a certain pre-buffering time, during which the playback buffer is filled 
by a determined amount of ordered chunks. The pre-buffering time is typically much shorter than the duration of the streaming session. 

The helpers may not have access to the whole video library, because of backhaul constraints or caching constraints.\footnote{For example, in a FemtoCaching network (see discussion in Section \ref{sec:intro-dilip}) each helper contains a subset of the files depending on some caching algorithm.} 
In general,  we denote by $\Hc(f)$ the set of helpers that 
contain file $f \in \Fc$.  Hence, user $u$ requesting file $f_u$ can only download video chunks from 
helpers in the set $\Nc(u) \cap \Hc(f_u)$. 

Each file $f \in \Fc$ is encoded at a finite number of different quality levels $m \in \{1, \ldots, N_f\}$. 
This is similar to the implementation of several current video streaming 
technologies, such as Microsoft Smooth Streaming and Apple HTTP Live Streaming~\cite{begen2011watching}. 
Due to the VBR nature of video coding \cite{ortega2000variable}, the quality-rate profile of a given file $f$ may vary from chunk to chunk. 
We let $D_f(m,t)$ and $B_f(m,t)$ denote the video quality measure (e.g., see \cite{wang2004image}) and the number of bits per pixel  
for file $f$ at chunk time $t$ and quality level  $m$ respectively. 

A scheduling policy for the network at hand consists of a sequence of decisions such that, at each chunk time $t$, 
each streaming user $u$ requests its desired $t$-th chunk of file $f_u$ from one or more helpers in $\Nc(u) \cap \Hc(f_u)$ at some
quality level $m_u(t) \in \{1, \ldots, N_f\}$, and  each helper $h$ transmits the source-encoded bits of currently or previously requested chunks 
to the users.  For simplicity, we assume that the scheduler time-scale coincides with the chunk interval, i.e., at each chunk interval 
a scheduling decision is made. Conventionally, we assume a slotted time axis $t = 0,1,2,3 \ldots,$ 
corresponding to epochs $t \times T_{\rm gop}$.  Letting $T_u$ denote the pre-buffering time of user $u$ (where $T_u$ is an integer), 
the chunks are downloaded starting at time $t = 0$ and the $t$-th chunk playback deadline is $t + T_u$. 
A buffer underrun event for user $u$ at time $t$ is defined as the event that the playback buffer 
does not contain chunk number $t - T_u$ at slot time $t$. When a buffer underrun even occurs, the playback may be stopped until
enough ordered chunks are accumulated in the playback buffer. This is called {\em stall event}, and the process of reconstituting the playback buffer
to a certain desired level of ordered chunks is referred to as {\em re-buffering}. Alternatively, the playback might just skip the missing chunk.
The details relative to pre-buffering, re-buffering and chunk skipping are discussed in  Section \ref{sec:prebuffering}.

Letting  $N_{\mathrm{pix}}$ denote the number of pixels per frame, 
a chunk contains $k = \eta T_{\rm gop} N_{\mathrm{pix}}$ pixels. 
Hence, the number of bits in the $t$-th chunk of file $f$, encoded at quality level $m$, is given by $k B_f(m,t)$. 
We assume that a chunk can be partially downloaded from multiple helpers, and
let $R_{hu}(t)$ denote the source coding rate (bit per pixel) of chunk $t$ requested by user $u$ from helper $h$. It follows that 
the source coding rates must satisfy, for all $t$, 
\begin{align}
\label{downloadconst}
\sum_{h \in \Nc(u) \cap \Hc(f_u)} R_{hu}(t) = B_{f_u}(m_u(t),t), \;\;\;\; \forall~(h, u) \in \Ec,
\end{align}
where $m_u(t)$ denotes the quality level at which chunk $t$ of file $f_u$ is requested by user $u$. 
The constraint (\ref{downloadconst}) reflects the fact that the aggregate bits of a given chunk $t$ from all helpers serving user $u$ must be equal to the 
total number of bits in the requested chunk. When a chunk request is made and the source coding rates 
$R_{hu}(t)$ are determined, helper $h$ places the corresponding $k R_{hu}(t)$ bits  in a transmission queue $Q_{hu}$ ``pointing'' at user $u$. 
This queue contains the source-encoded bits that have to be sent from helper $h$ to user $u$. 
Notice that in order to be able to download different parts of the same chunk from different helpers,
the network controller needs to ensure that all received bits from the serving helpers $\Nc(u) \cap \Hc(f_u)$ are useful, i.e., the union of all requested 
bits yields the total bits in the requested chunk, without overlaps or gaps. 
Alternatively, each chunk can be encoded by intra-session {\em Random Linear Network Coding}~\cite{ho2006random}  such that 
as long as $kB_{f_u}(m_u(t),t)$ parity bits are collected at user $u$, the $t$-th chunk  can be decoded and it becomes available in 
the user playback buffer.  
Interestingly, even though we optimize over algorithms that allow the possibility of downloading different bits of the same chunk from different helpers, 
the optimal scheduling policy (derived in Section~\ref{sec:numiidstate-dilip}) has a simple structure that always requests entire chunks from single helpers. 
Hence, without loss of optimality, neither protocol coordination to prevent overlaps or gaps, nor intra-session linear network coding, 
are needed for the algorithm implementation.
%

\subsection{Wireless transmission channel}  \label{phy-model}

We model the wireless channel for each link $(h,u) \in \Ec$ as a 
frequency and time selective underspread fading channel \cite{tse-viswanath}.
Using OFDM, the channel can be converted into a set of 
parallel narrowband sub-channels in the frequency domain (subcarriers), each of which is time-selective 
with a certain fading channel coherence time. The small-scale Rayleigh fading channel coefficients 
can be considered as constant over time-frequency ``tiles'' spanning blocks of adjacent subcarriers in the frequency domain and blocks of OFDM symbols 
in the time domain.  For example, in the LTE standard \cite{sesia-LTE}, the small scale fading coefficients can be considered
constant over a coherence time interval of $0.5$ ms and a coherence bandwidth of 180 kHz, corresponding to ``tiles''
of 7 OFDM symbols $\times$ 12 subcarriers.  For a total system available bandwidth of $18\mathrm{MHz}$
(after excluding the guard bands) and a scheduling slot of duration $T_{\rm gop} = 0.5$s (typical video chunk duration), 
we have that a scheduling slot spans $\frac{0.5 \times 18 \cdot10^{6}}{0.5 \cdot 10^{-3} \times 180 \cdot 10^3} = 10^5$ tiles, i.e., channel fading coefficients. 
Even assuming some correlation between fading coefficients, it is apparent that the time-frequency diversity experienced in the transmission of a 
chunk is {\em very large}.  Thus, it is safe to assume that channel coding over such a large number of resource blocks achieves the {\em ergodic capacity} 
of the underlying fading channel.

In this paper we refer to {\em ergodic capacity} as the {\em average} mutual information resulting from Gaussian i.i.d.
inputs of the single-user channel from helper $h$ and user $u \in \Nc(h)$, while treating the inter-cell interference, i.e., 
the signals of all other helpers $h'\neq h$ as noise,  where averaging is with respect to the first-order distribution of the small-scale fading.
This rate is achievable by i.i.d. Gaussian coding ensembles and approachable in practice by modern graph-based codes \cite{urbanke-book} 
provided that the length of a codeword spans a large number of independent small-scale fading states \cite{biglieri1998fading}.

We assume that the helpers transmit at constant power, and that the small-cell network makes use of universal frequency reuse, that is, 
the whole system bandwidth is used by all the helper stations.
We further assume that every user $u$, when decoding a  transmission from a particular helper $h \in \Nc(u)$ treats 
inter-cell interference as noise.  Under these system assumptions,  the maximum achievable rate\footnote{
We express channel coding rates  $\mu_{hu}$ in bit/s/Hz, i.e., bit per complex channel symbol use and the source coding rates $R_{hu}$ in bit/pixel, i.e., bits per source symbol, in agreement  with standard information theoretic channel coding and source coding.}
at slot time $t$ for link $(h,u) \in \Ec$ is given by 
\begin{equation}  \label{rateconst2}
C_{hu}(t) = \E{ \log\left(1+\frac{P_hg_{hu}(t)|s_{hu}|^2}{1+\sum_{\underset{h' \in \Nc(u)}{h' \neq h}} P_{h^{'}}g_{h^{'}u}(t)|s_{h^{'}u}|^2}\right)},
\end{equation} 
where $P_h$ is the transmit power of helper $h$, $s_{hu}$ is the small-scale fading gain from 
helper $h$ to user $u$ and $g_{hu}(t)$ is the slow fading 
gain (path loss) from helper $h$ to user $u$. Notice that at the denominator of the Signal to Interference plus Noise Ratio (SINR) inside the 
logarithm in (\ref{rateconst2}) we have the sum of the signal powers of all helpers $h' \in \Nc(u) : h' \neq h$, indicating the inter-cell interference suffered from 
user $u$, when decoding the transmission from helper $h$. 

In this work, consistently with most current wireless standards, we consider the case of intra-cell orthogonal access. 
This means that  each helper $h$ serves its neighboring users $u \in \Nc(h)$ using orthogonal FDMA/TDMA. 
It follows that the feasible set of channel coding rates $\{\mu_{hu}(t) : u \in \Nc(h)\}$ 
for each helper $h$ must satisfy the constraint:
\begin{equation}  \label{rateconst1}
\sum_{u \in \Nc(h)} \frac{\mu_{hu}(t)}{C_{hu}(t)} \leq 1,  \;\;\;\; \forall~h\in \Hc.
\end{equation}
The underlying assumption, which makes the rate region defined in (\ref{rateconst1}) achievable, is that helper $h$ is aware of 
the slowly varying path loss coefficients $g_{hu}(t)$ for all $u \in \Nc(h)$, such that rate adaptation is possible. This is 
consistent with currently implemented rate adaptation schemes \cite{sesia-LTE,802.11ac,molisch2010wireless}.

\subsection{Transmission queues dynamics and network state}

The dynamics (time evolution) of the transmission queues at the helpers is given by:
\begin{align}
Q_{hu}(t+1)=\max\{Q_{hu}(t) - n\mu_{hu}(t),0\}&+kR_{hu}(t),  \notag \\
& \forall~ (h,u) \in \Ec, \label{q-update}
\end{align}
where $n$ denotes the number of physical layer channel symbols corresponding to the duration 
$T_{\rm gop}$, and $\mu_{hu}(t)$ is the channel coding rate (bits/channel symbol) of the 
transmission from helper $h$ to user $u$ at time $t$. Notice that (\ref{q-update}) reflects the fact that at any chunk time $t$
the requested amount $kR_{hu}(t)$ of source-encoded bits is input to the queue of helper $h$ serving user $u$, 
and up to $n\mu_{hu}(t)$ source-encoded bits are extracted from the same queue and delivered by helper $h$ to user $u$ over the 
wireless channel.

The channel coefficients $g_{hu}(t)$ models  path loss and shadowing between helper $h$ and user $u$,
and are assumed to change slowly in time. For a typical small-cell scenario with nomadic users moving at walking speed or slower,
the path loss coefficients change on a time-scale of the order of  $10$s (i.e., $\approx 20$ scheduling slots).  
This time scale is much slower than the coherence of the small-scale fading, but it is comparable with the duration of the 
video chunks. Therefore, variations of these coefficients during a streaming session (e.g., due to user mobility) are relevant.  
At this point, we can formally define the network state and a feasible scheduling policy for our system.

\begin{defn} \label{network-state-def}
The network state $\omegav(t)$ collects the quantities that evolve independently of the scheduling decisions in the network. 
These are, in particular,  the slowly-varying channel gains, the video quality levels, 
and the corresponding bit-rates of the chunk at time $t$. Hence,  we have 
\begin{equation} \label{network-state} 
\omegav(t) = \left \{ g_{hu}(t), D_{f_u}(\cdot, t), B_{f_u}(\cdot, t) : \forall \; (h,u) \in \Ec \right \}. 
\end{equation}
\hfill $\lozenge$
\end{defn}

\begin{defn} \label{scheduling-policy}
A scheduling policy $\{a(t)\}_{t=0}^{\infty}$ is a sequence of control actions $a(t)$ comprising
the vector $\Rm(t)$ with elements $k R_{hu}(t)$ of requested source-coded bits, 
the vector $\muv(t)$ with elements $n\mu_{hu}(t)$ of transmitted channel-coded bits, 
and the quality level decisions $\{m_u(t) : \forall~u \in \Uc\}$. \hfill $\lozenge$
\end{defn}  

\begin{defn} \label{feasible-set}
For any $t$, the feasible set of control actions $A_{\omegav(t)}$ includes all control actions $a(t)$ such that
the constraints (\ref{downloadconst}) and (\ref{rateconst1}) are satisfied. \hfill $\lozenge$
\end{defn}

\begin{defn} \label{feasible-policy}
A feasible scheduling policy for the system at hand is a sequence of control actions  $\{a(t)\}_{t=0}^{\infty}$ such that
$a(t) \in A_{\omegav(t)}$ for all $t$. 
\hfill $\lozenge$
\end{defn}

\section{Problem Formulation and Optimal Scheduling Policy} \label{sec:numiidstate-dilip}

The goal of a scheduling policy for our system is to maximize a concave 
{\em network utility function} of the individual users' video quality index. 
Since these are time-varying quantities, we focus on the time-averaged expectation of such quantities. 
In addition, all source-coded bits requested by the users should be delivered. 
This imposes the constraint that all transmission queues at the helpers must be stable. 
Throughout this work, we use the following standard notation for the time-averaged expectation of any quantity $x$:
\begin{align}  \label{notation-bar}
\overline{x} := \lim_{t\rightarrow \infty}\frac{1}{t}\sum_{\tau=0}^{t-1}\E{x(\tau)}.
\end{align}
We define $\overline{D}_u:=\lim_{t\rightarrow \infty}\frac{1}{t}\sum_{\tau=0}^{t-1}\E{ D_{f_u}\left(m_u(\tau),\tau\right)}$ to be
the time-averaged expected quality of user $u$, and  $\overline{Q}_{hu} := \lim_{t\rightarrow \infty}\frac{1}{t}\sum_{\tau=0}^{t-1} \E{Q_{hu}\left(\tau\right)}$ to be the time-averaged expected length of the queue at helper $h$ for data transmission to user $u$, 
assuming temporarily that these limits exist.\footnote{The existence of these limits is assumed temporarily for ease of exposition of the 
optimization problem (\ref{NUMproblem}) but is not required for the derivation of the scheduling policy and for
the proof of Theorem \ref{main-result}.}
Let $\phi_u(\cdot)$ be a concave, continuous, and non-decreasing function 
defining network utility vs. video quality for user $u \in \Uc$. Then, the proposed scheduling policy is the solution of the following 
NUM problem:
\begin{align}
\textrm{maximize}  & \;\;\; \sum_{u \in \Uc}\phi_u(\overline{D}_u) \nonumber \\
 \textrm{subject to} & \;\;\; \overline{Q}_{hu} < \infty~\forall~ (h,u) \in \Ec  \nonumber \\
& \;\;\; a(t) \in A_{\omegav(t)}~\forall~t, \label{NUMproblem}
\end{align}
where requirement of finite $\overline{Q}_{hu}$ corresponds to the {\it strong stability} condition for all 
the queues \cite{neely2010stochastic}.

By appropriately choosing the functions $\phi_u(\cdot)$, we can impose some desired notion of fairness.
For example, a general class of concave functions suitable for this purpose is given by the $\alpha$-fairness network utility, 
defined by~\cite{mo2000fair} 
\begin{equation}
\phi_u(x) = \left \{ \begin{array}{ll}
\log x & \alpha = 1 \\
\frac{x^{1- \alpha}}{1 - \alpha} & \alpha > 0, \;\; \alpha \neq 1 \end{array} \right .
\end{equation} 
In this case, it is well-known that $\alpha = 0$ yields the maximization of the sum quality (no fairness), 
$\alpha \rightarrow \infty$ yields the maximization of the worst-case quality (max-min fairness) and
$\alpha = 1$ yields the maximization of the geometric mean quality (proportional fairness).

\begin{remark} \label{rem-smooth-streaming}
{\em On the relevance of NUM problem (\ref{NUMproblem}) for video streaming.}
A natural objection to our problem formulation is that queue stability guarantees only that chunks ``will be eventually delivered'' 
with finite (average) delay, but it does not guarantee that the chunks are delivered within their playback deadline.
In fact, the network utility function in (\ref{NUMproblem}) is defined in terms of the long-term averaged {\em requested} user video 
quality level. As a matter of fact, some requested chunks may not be delivered within their playback deadlines 
and therefore the {\em requested} video quality may not correspond to the {\em delivered} video quality. 
Of course, requested and delivered video quality coincide if the maximum delay incurred by any chunk 
requested by each user $u$ is not larger than the pre-buffering time $T_u$ allowed at the start-up phase of the streaming session of user $u$. 
As explained in Section \ref{sec:intro-dilip}, in order to obtain a clean and tractable problem leading to a low complexity and low-overhead 
decentralized policy, we have taken a ``divide and conquer'' approach. First, we focus on the NUM (\ref{NUMproblem}) subject to queue stability.
Then, we force the system to work in the smooth streaming regime by allowing sufficient 
pre-buffering time. This is obtained by the decentralized adaptive pre-buffering/re-buffering time 
estimation scheme presented in Section \ref{sec:prebuffering}. 

Here, we argue that the proposed scheme can achieve near-optimal performance in the sense of a small perturbation with respect to
the optimality of a modified network utility function where the buffer underrun events are weighted by some {\em bounded} 
penalty in terms of the quality index. 

First, we observe that if, for each $u$, the delay introduced by all queues in the helpers serving 
$u$ is upperbounded by a deterministic constant $E_{u,\max}$, then by letting the pre-buffering 
time $T_u \geq E_{u,\max}$  all chunks requested at time $t$ are delivered within their deadline $t - T_u$. 
In this case,  the buffer underrun rate is zero and our policy (solution of
the NUM  problem (\ref{NUMproblem})) is {\em exactly} optimal even with respect to a modified network utility function that takes into account
the buffer underrun events.  

Then, we observe that, in the realistic case of non-stationary non-ergodic networks considered in this paper, 
such uniform delay upper bounds may not exist or may be simply too loose  to yield a practically useful pre-bufffering policy. 
For this purpose, the adaptive pre-buffering/re-buffering policy proposed in Section \ref{sec:prebuffering} provides the best  possible
estimate of $T_u$ based on local information, such that the buffer underrun rate can be made small. 
Define the indicator function\footnote{$1\{\Ac\}$ denotes the indicator function of a  condition or event $\Ac$.} 
of the buffer underrun event for user $u$ at time $t$ as 
$\epsilon_u(t) = 1\{\mbox{chunk $t$ is not delivered by time $t + T_u$}\}$ and  let $\overline{\epsilon}_u$ denote its
time-averaged expected value (according to the notation defined in (\ref{notation-bar})), i.e., the buffer underrun rate of user $u$.
Let also $0 \leq \Xi_u(t) \leq \Xi_u$ denote the video quality penalty incurred by such event, assumed 
uniformly bounded by the user-dependent constant $\Xi_u$. 
The modified network utility function that takes explicitly into account the buffer underrun events
is given by $\sum_{u \in \Uc} \phi_u(\overline{D}_u - \overline{\Xi}_u)$. 
Since $\phi_u(\cdot)$ is concave and non-decreasing (it has positive bounded variation), we can write
\begin{equation} 
0 \leq \frac{\phi_u(\overline{D}_u) - \phi_u(\overline{D}_u - \overline{\epsilon}_u \Xi_u)}{\overline{\epsilon}_u \Xi_u} \leq \phi'_u,
\end{equation}
for some positive constant $\phi'_u$. Summing over all $u \in \Uc$ and using the fact that 
$\overline{\Xi}_u \leq \overline{\epsilon}_u \Xi_u$,  which implies 
$\phi_u(\overline{D}_u - \overline{\Xi}_u) \geq \phi_u(\overline{D}_u - \overline{\epsilon}_u \Xi_u)$, 
we obtain the bounds
\begin{equation}  
\sum_{u \in \Uc} \phi_u(\overline{D}_u) \geq \sum_{u \in \Uc} \phi_u(\overline{D}_u - \overline{\Xi}_u) 
\geq \sum_{u \in \Uc} \phi_u(\overline{D}_u) - \sum_{u\in \Uc} \phi'_u \overline{\epsilon}_u \Xi_u. 
\end{equation}
Hence, the maximization in (\ref{NUMproblem}), combined with
a pre-buffering/re-buffering scheme that makes
the buffer underrun rate  $\overline{\epsilon}_u$ very small for all users, 
yields a modified network utility function
(including the quality penalty incurred by buffer underrun events) within a small perturbation of the optimal value of (\ref{NUMproblem}).
The latter clearly upper bounds any policy that takes explicitly into account the chunk delivery delays, 
since it is given in terms of the {\em requested} video quality. 
In conclusions, when ``almost all'' chunks are delivered within their playback time, 
maximizing the network utility expressed in terms of the requested video quality is {\em nearly optimal} and, 
as shown in Sections \ref{our-scheduling-policy} -- \ref{optimality},
has the advantage of yielding a very simple decentralized dynamic scheduling policy 
through the DPP approach. 
\hfill $\lozenge$

\end{remark}

Having clarified that the solution of the NUM problem (\ref{NUMproblem}) is relevant for the VoD streaming problem at hand, 
in the following we first illustrate a dynamic scheduling policy for problem (\ref{NUMproblem}) and then 
provide Theorem \ref{main-result}, which states the optimality guarantee  of the proposed dynamic scheduling policy 
in a strong per-sample path sense.

\subsection{Dynamic scheduling policy}  \label{our-scheduling-policy}

We introduce auxiliary variables $\gamma_u(t)$ and corresponding virtual queues  
$\Theta_u(t)$ with buffer evolution: 
\begin{align}
\Theta_u(t+1) = \max{\{\Theta_u(t) + \gamma_u(t) - D_{f_u}(m_u(t),t),0 \}}.  \label{virt-update}
\end{align} 
Each user $u \in \Uc$ updates its own virtual queue $\Theta_u(t)$ locally. Also, we introduce a scheduling policy control parameter $V > 0$ that trades off
the average queue lengths with the accuracy with which the policy is able to approach the optimum of the NUM problem (\ref{NUMproblem}). 

According to Definition \ref{scheduling-policy}, a scheduling policy is defined by specifying how to calculate the source-coding rates 
$R_{hu}(t)$, the video quality levels $m_u(t)$, and the channel coding rates $\mu_{hu}(t)$,  
for all chunk times $t$. These are given by solving local maximizations at each user node $u$ and helper node $h$. 
Since these maximizations depend only on local variables that can be learned 
by each node from its neighbors through simple protocol signaling at negligible overhead cost (a few scalar quanties 
per chunk time),  the resulting policy is decentralized.

\subsubsection{Control actions at the user nodes (congestion control)}

At time $t$, each $u \in \Uc$ chooses the helper in its neighborhood having the desired 
file $f_u$ and with the shortest queue, i.e., 
\begin{equation} \label{helper-selection}
h^*_u(t)  = \mbox{argmin} \left \{  Q_{hu}(t) \; : \;  h \in \Nc(u)\cap \Hc(f_u) \right \}. 
\end{equation}
Then, it determines the quality level $m_u(t)$ of the requested chunk at time $t$ as:
\begin{align} 
m_u(t) = \mbox{argmin} \big \{ k & Q_{h^*_u(t) u}(t) B_{f_u}(m,t) - \Theta_u(t) D_{f_u}(m,t) \notag \\ 
& : m \in \{1, \ldots, N_{f_u}\} \big \}. 
 \label{quality-level-decision}
\end{align}
The source coding rates for the requested chunk at time $t$ are given by: 
\begin{equation} \label{source-coding-decision}
R_{hu}(t) = \left \{ \begin{array}{ll}
B_{f_u}(m_u(t),t) & \mbox{for} \; h = h^*_u(t) \\
0  & \mbox{for} \; h \neq h^*_u(t)
\end{array} \right .
\end{equation}
The virtual queue $\Theta_u(t)$ is updated according to (\ref{virt-update}), where 
$\gamma_u(t)$ is given by:
\begin{equation} \label{opt-gamma}
\gamma_u(t) = \mbox{argmax} \left \{  V\phi_u(\gamma) - \Theta_u(t) \gamma \; : \;  \gamma \in [D_u^{\min}, D_u^{\max}] \right \},
\end{equation}
where $D_u^{\min}$ and $D_u^{\max}$ are uniform lower and upper bounds on the quality 
function $D_{f_u}(\cdot, t)$.

We refer to the policy (\ref{helper-selection}) -- (\ref{opt-gamma}) as {\em congestion control} since 
each user $u$ selects the helper from which to request the current video chunk and the quality at which 
this chunk is requested by taking into account the state of the transmission queues of all helpers $h$ that potentially can deliver such chunk, 
and choosing the least congested queue (selection in (\ref{helper-selection})) and an appropriate video quality level that balances 
the desire for high quality (reflected by the term $-\Theta_u(t) D_{f_u}(m,t)$ in (\ref{quality-level-decision})) and 
the desire for low transmission queues (reflected by the term  $k Q_{h^*_u(t) u}(t) B_{f_u}(m,t)$ in (\ref{quality-level-decision})). 
Notice that the streaming of the video file $f_u$ may be handled by {\em different} helpers across
the streaming session, but each individual chunk is entirely downloaded from a single helper. 
Notice also that in order to compute (\ref{helper-selection}) -- (\ref{opt-gamma}) 
each user needs to know only {\em local information} formed by the queue backlogs $Q_{hu}(t)$ of its neighboring helpers, 
and by the locally computed virtual queue backlog $\Theta_u(t)$. 

The above congestion control action at the users is reminiscent of the current  adaptive streaming technology for video on demand systems, 
referred to as DASH (Dynamic Adaptive Streaming over HTTP)~\cite{sanchez2011improved,sanchez2011idash}, 
where the client (user) progressively fetches a video file by downloading successive chunks, 
and makes adaptive decisions on the quality level based on its current knowledge of the congestion of the 
underlying server-client connection. Our policy generalizes DASH by allowing the client $u$ to dynamically select the
least backlogged server $h^*_u(t)$, at each chunk time $t$.

\subsubsection{Control actions at the helper nodes (transmission scheduling)}  \label{rate-scheduling-sec}

{ At time $t$, the general transmission scheduling consists of maximizing the weighted sum rate of the transmission rates achievable at scheduling slot $t$.
Namely, the network of helpers must solve the Max-Weighted Sum Rate (MWSR) problem:
\begin{align} \label{mwsr-general}
\mbox{maximize} & \;\;\; \sum_{h \in \Hc} \sum_{u \in \Nc(h)} Q_{hu}(t) \mu_{hu}(t) \nonumber \\
\mbox{subject to} & \;\;\; \muv(t) \in \Rc(t) 
\end{align}
where $\Rc(t)$ is the region of achievable rates supported by the network at time $t$. 
In this work, we consider two different physical layer assumptions, yielding to two versions of the above general MWSR problem.

In the first case, referred to as ``macro-diversity'', the users can decode multiple data streams from multiple helpers
if they are scheduled with non-zero rate on the same slot. Notice that, consistently with (\ref{rateconst2}) and (\ref{rateconst1}), 
this does not contradict the fact that interference is treated as noise and that each helper uses orthogonal intra-cell access.\footnote{As a matter of fact, 
if a user is scheduled with non-zero rate at more than one helper, it could use successive interference cancellation or joint decoding of
all its intended data streams. Nevertheless, we assume here, conservatively, that interference (even from the intended streams)
is treated as noise.}
In the macro-diversity case, the rate region $\Rc(t)$ is given by the Cartesian product of 
the orthogonal access regions (\ref{rateconst1}), such that the general MWSR problem (\ref{mwsr-general}) decomposes into
individual problems, to be solved in a decentralized way at each helper node. 
After the change of variables  $\nu_{hu}(t) = \frac{\mu_{hu}(t)}{C_{hu}(t)}$, 
it is immediate to see that (\ref{mwsr-general}) reduces to the set of decoupled {\em Linear Programs} (LPs):
\begin{eqnarray}  \label{sucasuca}
 \mbox{maximize} & & \sum_{u \in \Nc(h)} Q_{hu}(t)C_{hu}(t) \nu_{hu}(t) \nonumber \\
 \mbox{subject to} & & \sum_{u \in \Nc(h)} \nu_{hu}(t) \leq 1, 
\end{eqnarray}
for all $h \in \Hc$.  The feasible region of (\ref{sucasuca}) 
is the $|\Nc(h)|$-dimensional simplex  and the solution is given by the vertex corresponding to user $u^*_h(t)$ given by 
\begin{equation}  \label{user-selection}
u^*_h(t) = \mbox{argmax} \left \{ Q_{hu}(t) C_{hu}(t) \; : \; u \in \Nc(h) \right \}, 
\end{equation}
with rate vector given by $\mu_{h u^*_h(t)}(t) = C_{h u^*_h(t)}(t)$ and $\mu_{hu}(t) = 0$ for all $u \neq u^*_h(t)$. 

In the second case, referred to as ``unique association'', any user can receive data from not more than a single helper on any scheduling slot. 
In this case, the MWSR problem reduces to a maximum weighted matching problem that can be solved by an LP as follows.
We introduce variables $\alpha_{hu}(t)$ such that $\alpha_{hu}(t) = 1$ if user $u$ is served by helper $h$ at time $t$ and 
$\alpha_{hu}(t) = 0$ if it is not. It is obvious that if $\alpha_{hu}(t) = 1$, then $\mu_{hu}(t) = C_{hu}(t)$, implying that
$\mu_{h'u}(t) = 0$ for all $h' \neq h$ and $\mu_{hu'}(t) = 0$ for all $u' \neq u$ (by (\ref{rateconst1})). 
Hence, (\ref{mwsr-general}) in this case reduces to
\begin{align} \label{mwsr-matching}
\mbox{maximize} & \;\;\; \sum_{h \in \Hc} \sum_{u \in \Nc(h)} \alpha_{hu}(t) Q_{hu}(t) C_{hu}(t) \nonumber \\
\mbox{subject to} & \;\;\; \sum_{h \in \Nc(u)} \alpha_{hu}(t) \leq 1~\forall~u \in \Uc, \nonumber \\
& \;\;\; \sum_{u \in \Nc(h)} \alpha_{hu}(t) = 1~\forall~h \in \Hc, \nonumber\\
& \;\;\; \alpha_{hu}(t) \in \{0,1\}, \;\; \forall \;\; h \in \Hc, \; u \in \Uc.
\end{align}
A well-known result (see \cite[Theorem 64.7]{schrijver2003combinatorial}) states that, since the network graph 
$\Gc = (\Uc, \Hc, \Ec)$ is bipartite, 
the integer programming problem (\ref{mwsr-matching}) can be relaxed to an LP
by replacing the integer constraints on $\{\alpha_{hu}(t)\}$ with the linear constraints $\alpha_{hu}(t) \in [0,1]$ for all 
$h \in \Hc, \; u \in \Uc$. The solution of the relaxed LP is guaranteed to be integral, such that it is feasible
(and therefore optimal) for (\ref{mwsr-matching}).  
Notice that, in the case of unique association, the rate scheduling problem does not admit a 
decoupled solution, calculated independently at each helper node. Hence, a network controller that solves 
(\ref{mwsr-matching}) and allocates the downlink rates (and the user-helper dynamic association) at each slot time $t$
is required. Again, since $t$ ticks at the chunk time, i.e., on the time scale of seconds, 
this does not involve a very large complexity, although it is definitely more complex than the macro-diversity case. 

\begin{remark} \label{rem-dynamic-association}
{\em Dynamic helper-user association.}
Notice that here, unlike conventional cellular systems,  we do not assign a fixed set of users to 
each helper.  In contrast, the helper-user association is dynamic, and results from the transmission scheduling 
decision. Notice also that, for both the macro-diversity and the unique association cases, 
despite the fact  that each helper $h$ is allowed to serve its queues
with rates $\mu_{hu}(t)$ satisfying (\ref{rateconst1}), the proposed policy allocates the whole $t$-th 
downlink slot to a single user $u \in \Nc(h)$, served at its own peak-rate $C_{hu}(t)$. 
This is reminiscent of opportunistic user selection in high-rate downlink schemes 
of 3G cellular systems, such as HSDPA and Ev-Do \cite{sesia-LTE,bhushan2006cdma2000}. \hfill $\lozenge$
\end{remark}
}

\subsection{Derivation of the scheduling policy}  \label{derivation}

In order to solve problem  (\ref{NUMproblem}) using the stochastic optimization theory developed in \cite{neely2010stochastic}, it is
convenient to transform it into an equivalent problem that involves  the maximization of a single time average. 
This transformation is achieved through the use of auxiliary variables $\gamma_u(t)$ and the corresponding virtual queues 
$\Theta_u(t)$ with buffer evolution given in (\ref{virt-update}).  Consider the transformed problem:
\begin{align}
 \textrm{maximize} & \;\;\; \sum_{u \in \Uc}\overline{\phi_u({\gamma}_u)}\label{maxutiltrans}\\
 \textrm{subject to} & \;\;\;  \overline{Q}_{hu} < \infty~\forall~ (h,u) \in \Ec \label{qstableconsttrans}\\
& \;\;\; \overline{\gamma}_u \leq \overline{D}_u~\forall~u~\in~\Uc \label{gammaconst}\\
& \;\;\; D_u^{\min} \leq \gamma_u(t) \leq D_u^{\max}~\forall~u~\in~\Uc \label{rectconst}\\
& \;\;\; a(t) \in A_{\omegav(t)}~\forall~t \label{feasibleoptionstrans}
\end{align}
Notice that constraints~(\ref{gammaconst}) correspond to stability of the virtual queues $\Theta_u$, since
$\overline{\gamma}_u$ and $\overline{D}_u$ are the time-averaged arrival rate and the time-averaged 
service rate for the virtual queue given in (\ref{virt-update}).  We have: 

\begin{lemma} \label{equivalence}
Problems (\ref{NUMproblem}) and (\ref{maxutiltrans}) -- (\ref{feasibleoptionstrans}) are equivalent. 
\end{lemma}

\begin{IEEEproof} See Appendix \ref{proof-lem}. \end{IEEEproof}

Thanks to Lemma \ref{equivalence}, we shall now focus on the solution of problem (\ref{maxutiltrans}) -- (\ref{feasibleoptionstrans}).
Let $\Qm(t)$ denote the column vector containing the backlogs of queues $Q_{hu}~\forall~(h,u)\in\Ec$,  
let $\Thetam(t)$ denote the column vector for the virtual queues $\Theta_u~\forall~u\in\Uc$,
$\gammav(t)$ denote the column vector with elements $\gamma_u(t)~\forall~u\in\Uc$, and 
$\Dm(t)$ denote the column vector with elements $D_{f_u}(m_u(t),t)~\forall~u\in\Uc$. 
Let $\Gm(t) = \left[ \Qm^\transp(t), \Thetam^\transp(t) \right]^\transp$ be the composite vector of queue backlogs
and define the quadratic Lyapunov function $L(\Gm(t)) = \frac{1}{2} \Gm^\transp (t) \Gm(t)$.  The one-slot drift of the Lyapunov function at slot $t$ is given by
\begin{align}
\label{lyapbound}
 &L(\Gm(t+1))-L(\Gm(t)) \notag \\
 & = \frac{1}{2}\left (\Qm^\transp(t+1) \Qm(t+1)- \Qm^\transp (t) \Qm(t) \right ) \notag \\
&~~~~~~+ \frac{1}{2} \left ( \Thetam^\transp(t+1) \Thetam(t+1)- \Thetam^\transp (t) \Thetam(t) \right )\notag\\
& =  \frac{1}{2}\left[ \left( \max\{ \Qm(t) - \muv(t), {\bf 0} \} + \Rm(t) \right )^\transp \left(\max\{\Qm(t)- \muv(t), {\bf 0} \} \right. \right. \notag \\
&~~~~~~ \left. \left. +~\Rm(t)\right)- \Qm^\transp(t) \Qm(t) \right] \notag \\
&~~~~ + \frac{1}{2}\left[\left(\max\{ \Thetam(t)+\gammav(t)-\Dm(t), {\bf 0}\} \right)^\transp\left(\max\{\Thetam(t)+
\gammav(t) \right. \right. \notag \\
&~~~~~~~~\left. \left. - \Dm(t), {\bf 0} \} \right) - \Thetam^\transp(t)\Thetam(t)\right],
\end{align}

where we have used the queue evolution equations (\ref{q-update}) and (\ref{virt-update}) and ``max'' is applied componentwise.  
 
Noticing that for any non-negative scalar quantities $Q,\mu, R, \Theta,\gamma$ and $D$ we have the inequalities
\begin{align}
(\max\{Q-\mu, 0\}+R)^2 \leq Q^2+\mu^2+R^2+2Q(R-\mu), \label{ineq1}
\end{align}
and 
\begin{align}
(\max\{\Theta + \gamma - D, 0\})^2 &\leq (\Theta + \gamma -D)^2 \notag \\
& = \Theta^2 + (\gamma-D)^2 + 2\Theta(\gamma-D), \label{ineq2}
\end{align}
we have
\begin{align}
& L({\bf G}(t+1))-L({\bf G}(t)) \notag \\
&\leq  \frac{1}{2}{\boldsymbol \mu}^\transp(t){\boldsymbol \mu}(t)+{\bf R}^\transp(t){\bf R}(t)+\left({\bf R}(t)-{\boldsymbol \mu}(t)\right)^
\transp{\bf Q}(t) \notag \\
&~+ \frac{1}{2}\left(\gammav(t)-\Dm(t)\right)^\transp\left(\gammav(t)-\Dm(t)\right) + \left(\gammav(t)-
\Dm(t)\right)^\transp\Thetav(t) \label{drift-bound} \\
&\leq  \Kc + \left({\bf R}(t)-{\boldsymbol \mu}(t)\right)^\transp{\bf Q}(t)+\left(\gammav(t)-\Dm(t)\right)^\transp\Thetav(t),  \label{Kc}
\end{align}
where $\Kc$ is a uniform bound on the term $\frac{1}{2}\left[{\boldsymbol \mu}^\transp(t){\boldsymbol \mu}(t)+{\bf R}^\transp(t){\bf R}(t)\right]
+ \frac{1}{2}\left(\gammav(t)-\Dm(t)\right)^\transp\left(\gammav(t)-\Dm(t)\right)$, which exists under the 
realistic assumption that the source coding rates, the channel coding rates and the video quality measures are upper 
bounded by some constants, independent of $t$. The conditional expected {\it Lyapunov drift} for slot $t$ is defined by 
\begin{equation}
\label{drift-def}
 \Delta({\bf G}(t)) = \E{L({\bf G}(t+1))|{\bf G}(t)}-L({\bf G}(t)).
\end{equation}
Adding on both sides the penalty term $ -V \sum_{u \in \Uc} \E{\phi_u(\gamma_u(t))|{\bf G}(t)}$, where $V \geq 0$ is the policy control parameter
already introduced above, we have
\begin{align}
\label{dpp-ineq}
&\Delta({\bf G}(t))-V\sum_{u \in \Uc}\E{\phi_u(\gamma_u(t))|{\bf G}(t)}  \leq \Kc  \notag \\
& -V\sum_{u \in \Uc}\E{\phi_u(\gamma_u(t))|{\bf G}(t)}+\E{\left({\bf R}(t)-
\boldsymbol \mu(t)\right)^\transp{\bf Q}(t)|
{\bf G}(t)} \notag \\
&~~~ +\E{\left({\gammav}(t)-
\Dm(t)\right)^\transp{\Thetav}(t)|
{\bf G}(t)}.
\end{align}
The DPP policy acquires information about ${\bf G}(t)$ and $\omegav(t)$ at every slot $t$ and chooses 
$a(t)  \in A_{\omegav(t)}$ in order to minimize the right hand side of the above inequality. 
The non-constant part of this expression can be written as
\begin{align}
\left [ {\bf R}^\transp (t) {\bf Q}(t) - \Dm^\transp(t)\Thetav(t)\right ]  
 &- \left [  V\sum_{u \in \Uc}\phi_u(\gamma_u(t)) -  \gammav^\transp(t)\Thetav(t)  \right ] \notag \\
&- {\boldsymbol \mu}^\transp (t){\bf Q}(t).
\label{DPP}
\end{align}
The resulting control action $a(t)$ is given by the minimization, at each chunk time $t$, of the expression in (\ref{DPP}). 
Notice that the first term of (\ref{DPP}) depends only on $\Rm(t)$ and on $m_u(t)~\forall~u \in \Uc$, 
the second term of (\ref{DPP}) depends only on $\gammav(t)$ and 
the third term of (\ref{DPP}) depends only on $\muv(t)$.
Thus, the overall minimization decomposes into three separate sub-problems. 
The first sub-problem (related to the first term in (\ref{DPP})) consists of choosing the quality levels $\{m_u(t)\}$ 
and the requested video-coding rates $\{R_{hu}(t)\}$ for each user $u$ and current chunk at time $t$. 
The second sub-problem (related to the second term in (\ref{DPP})) involves the greedy maximization of each user network utility function 
with respect to the auxiliary control variables $\gamma_u(t)$.  The third sub-problem (related to the third term in (\ref{DPP})), 
consists of allocating the channel coding rates $\mu_{hu}(t)$ for each helper $h$ to its neighboring users $u \in \Nc(h)$. 

Next, we show that the minimization of (\ref{DPP}) yields the congestion control sub-policy at the users and the transmission scheduling
sub-policy at the helpers given before. 

\subsubsection{Derivation of the congestion control action}
The first term in (\ref{DPP}) is given by 
\begin{align}
\sum_{u \in \Uc} \left \{ \sum_{h \in \Nc(u) \cap \Hc(f_u)} k Q_{hu}(t)R_{hu}(t) - \Theta_u(t) D_{f_u}\left(m_u(t),t\right) \right \}.
\end{align}
The minimization is achieved by minimizing separately each term inside the sum w.r.t. $u$ with respect to
$m_u(t)$ and $R_{hu}(t)$.  It is immediate to see that the solution consists of choosing the helper $h^*_u(t)$ as in (\ref{helper-selection}), 
the quality level as in (\ref{quality-level-decision}) and requesting the whole chunk from helper $h^*_u(t)$
at quality $m_u(t)$, i.e., letting $R_{h^*_u(t) u}(t) = B_{f_u}(m_u(t),t)$, as given in (\ref{source-coding-decision}). 
The second term in (\ref{DPP}), after a change of sign, is given by 
\begin{align}
\sum_{u \in \Uc} \left \{ V \phi_u(\gamma_u(t)) -  \gamma_u(t) \Theta_u(t) \right \}. 
\end{align}
Again, this is maximized by maximizing separately each term, yielding (\ref{opt-gamma}). 

\subsubsection{Derivation of the transmission scheduling action} 

After a change of sign, the maximization of the third term in (\ref{DPP}) yields precisely
(\ref{mwsr-general}) where $\Rc(t)$ is defined by the physical layer model of the network, 
and it is particularized to the cases of macro-diversity and unique association as discussed in Section \ref{rate-scheduling-sec}.

{ It is worthwhile to notice here that our NUM approach can be applied to virtually any network with any physical layer
(e.g., including non-universal frequency reuse, non-orthogonal intra-cel access, 
multiuser MIMO \cite{caire2010multiuser}, cooperative network MIMO \cite{Huh11}). In fact, all what is needed is to characterize the network
in terms of its achievable rate region $\Rc(t)$, when averaging with respect to the small-scale fading, and 
conditioning with respect to the slowly time-varying pathloss coefficients, that depend on the network topology and therefore on
the users motion. Of course, for more complicated type of wireless physical layers, the description of $\Rc(t)$ and therefore the solution of
the corresponding MWSR problem (\ref{mwsr-general}) may be much more involved than in the cases treated here. 
For example, an extension of this approach to the case of multi-antenna helper nodes using multiuser MIMO
is given in \cite{bethanabhotla2014adaptive}.} 

\subsection{Optimality}  \label{optimality}

As outlined in Section~\ref{sec:sysmodel-dilip}, VBR video yields time-varying quality and rate functions $D_f(m, t)$ and $B_f(m, t)$, which 
depend on the individual video file. Furthermore, arbitrary user motion yields time variations of the path coefficients $g_{hu}(t)$ at the same time-scale of the video streaming 
session. As a result, any stationarity or ergodicity assumption about the network state process $\omegav(t)$ is unlikely to hold in most practically 
relevant settings. Therefore, we consider the optimality of the DPP policy for an {\em arbitrary sample path} of the network state $\omegav(t)$. 
Following in the footsteps of \cite{neely2010stochastic,neely2010universal}, we compare the network utility achieved by our DPP policy  with that
achieved by an optimal oracle policy with  $T$-slot lookahead, i.e., such knowledge of the future network states over an interval of length $T$ slots. 
Time is split into frames of duration $T$ slots and we consider $F$ such frames. For an arbitrary sample path $\omegav(t)$, we consider the static 
optimization problem over the $j$-th frame
\begin{align}
\mbox{maximize} &\;\;\sum_{u \in \Uc} \phi_u\left(\frac{1}{T}\sum_{\tau=jT}^{(j+1)T-1} D_u(\tau)\right)   \label{obj-arbit}\\
\mbox{subject to} &\;\;\; \frac{1}{T} \sum_{\tau=jT}^{(j+1)T-1}\left[ kR_{hu}\left(\tau\right)-n\mu_{hu}\left(\tau\right)\right]\leq0 \notag \\ 
&~~~~~~~~~~~~~~~~~~~~~~~~~~~~~~~~~~\forall~(h, u) \in \Ec \label{arbit-const} \\
& \;\;\; a(t) \in A_{\omegav(t)}~\forall~t~\in~\{jT, \ldots, (j+1)T-1\}, \label{feas-action-arbit} 
\end{align}
and denote by $\phi_j^{\rm opt}$ the maximum of the network utility function for 
frame $j$,  achieved over all policies which have future knowledge of the sample 
path $\omegav(t)$ over the  $j$-th frame, subject to the constraint (\ref{arbit-const}), 
which ensures that for every queue $Q_{hu}$, the total service provided 
over the frame is at  least as large as the total arrivals in that frame.  We have the following result:

\begin{thm} \label{main-result}
For the system defined in Section \ref{sec:sysmodel-dilip}, with state, scheduling policy and feasible action set
given in Definitions \ref{network-state}, \ref{scheduling-policy} and \ref{feasible-set}, respectively, the dynamic scheduling policy 
defined in Section  \ref{our-scheduling-policy}, with control actions given in (\ref{virt-update}) -- (\ref{user-selection}), 
achieves the per-sample path network utility
\begin{align}
\sum_{u \in \Uc} \phi_u \left( \overline{D}_u \right) \geq 
\lim_{F \rightarrow \infty} \frac{1}{F}\sum_{j=0}^{F-1}\phi_j^{\rm opt} -  O\left (\frac{1}{V} \right ) \label{optutil-arbit1}
\end{align}
with bounded queue backlogs satisfying
\begin{align}
\lim_{F \rightarrow \infty} \frac{1}{FT} \sum_{\tau=0}^{FT-1}\left( \sum_{(h,u) \in \Ec}Q_{hu}(\tau) 
+ \sum_{u \in \Uc}\Theta_{u}(\tau)\right) \leq  O(V) \label{strongstab-arbit1}
\end{align}
where $O(1/V)$ indicates a term that vanishes as $1/V$ and $O(V)$ indicates a term that grows linearly with $V$, as the policy control parameter 
$V$ grows large. 
\end{thm}

\begin{IEEEproof} See Appendix \ref{proof-thm}. \end{IEEEproof}

An immediate corollary of Theorem \ref{main-result} is:
\begin{cor}  \label{corcor}
For the system defined in Section \ref{sec:sysmodel-dilip}, when the network state is stationary and ergodic, then 
\begin{align}
\sum_{u \in \Uc} \phi_u(\overline{D}_u) \geq \phi^{\rm opt} - O\left ( \frac{1}{V}\right ), 
\label{utilperf-iid}
\end{align}
where $\phi^{\rm opt}$ is the optimal value of the NUM problem (\ref{NUMproblem}) in the stationary ergodic 
case,\footnote{Notice that in the stationary and ergodic case the value $\phi^{\rm opt}$ is generally achieved
by an instantaneous policy with perfect knowledge of the state statistics or, equivalently, by a policy with infinite look-ahead, since
the state statistics can be learned arbitrarily well from any sample path with probability 1, because of ergodicity.} 
and
\begin{align}
\sum_{(h,u) \in \Ec} \overline{Q}_{hu} + \sum_{u \in \Uc}\overline{\Theta}_u \leq  O(V)
\label{strongstab-iid}
\end{align}
In particular, if the network state is i.i.d., the bounding term in (\ref{utilperf-iid}) is explicitly given by 
$O(1/V) = \frac{\Kc}{V}$, and the bounding term in (\ref{strongstab-iid}) is explicitly given by 
$\frac{\Kc + V( \phi_{\max} - \phi_{\min} )}{\epsilon}$, where 
$\phi_{\min} = \sum_{u \in \Uc}\phi_u(D_u^{\min})$, $\phi_{\max} = \sum_{u \in \Uc}\phi_u(D_u^{\max})$, 
$\epsilon > 0$ is the slack variable corresponding to the constraint (\ref{arbit-const}), 
and the constant $\Kc$ is defined in (\ref{Kc}). 
\end{cor}

\begin{IEEEproof} See Appendix \ref{proof-thm}. \end{IEEEproof}

\section{Pre-buffering, re-buffering and skipping chunks}  \label{sec:prebuffering} 

As described in Section \ref{sec:sysmodel-dilip}, the playback process consumes chunks  at fixed playback rate 
$1/T_{\rm gop}$ (one chunk per time slot), while the number of {\em ordered} chunks per unit time entering the 
playback buffer is a random variable,  due to the fact that the network state $\omegav(t)$ is
a random process (or an arbitrary varying function of time) and the transmission resources are dynamically 
allocated by the scheduling policy. Chunks must be ordered sequentially in order to be useful for video playback. 
If chunks go through different queues in the network and are affected by different delays, it may happen that
already received chunks with higher order number cannot be used for playback until the missing chunks with lower order number 
are also received. 

{ As we have noticed already in Section \ref{sec:intro-dilip} and in Remark \ref{rem-smooth-streaming}, 
the NUM problem formulation in (\ref{NUMproblem}) does not take into account
the possibility of buffer underrun events, i.e., chunks that are not delivered within their playback deadline. 
This simplification has the advantage of yielding the simple and decentralized scheduling policy of 
Section \ref{our-scheduling-policy}. However, in order to make such policy useful in practice
we have to force the system to work in the smooth streaming regime, i.e., in the regime of very small buffer underrun rate. 
This can be done by adaptively determining the pre-buffering time $T_u$ for each user $u$ 
on the basis of an estimate of the largest delay of queues $\{Q_{hu}(t) : h \in \Nc(u)\}$. 
In this section, we propose a simple method that allows to determine $T_u$ in a decentralized way, 
based on the {\em local information} available at each user $u$.} 

\begin{figure*}
\centering
\includegraphics[width = 80 mm, height = 50mm]{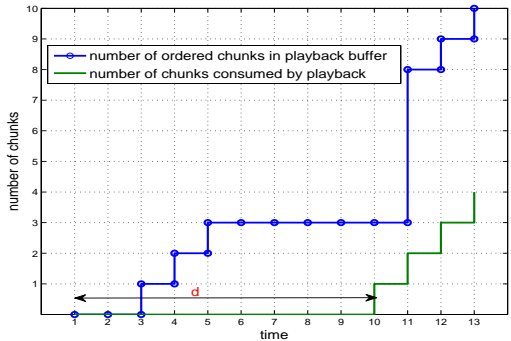}
\caption{Evolution of number of ordered and consumed chunks}
\label{chunkarriv-fig}
\end{figure*}

An example of the playback buffer dynamics is illustrated in Table~\ref{chunkarriv-tab} and Fig.~ \ref{chunkarriv-fig}. 
The table indicates the chunk numbers and their respective arrival times. 
The blue curve in Fig.~~\ref{chunkarriv-fig} shows the time evolution of the number of ordered chunks 
available in the playback buffer. The green curve indicates the evolution with time of the number of chunks consumed by playback. 
The playback consumption starts after an initial pre-buffering delay $T_u = d$, as indicated in the figure. 
At any instant $t$, the chunk requested at $t-d$ is expected to be available in the playback buffer. 
However, if the chunk is delivered with a delay greater than $d$, the two curves meet and a buffer underrun 
event occurs.  In order to prevent these events, each user $u$ should choose its pre-buffering time $T_u$ to be larger than 
the maximum delay of the serving queues $\{Q_{hu} : h \in \Nc(u) \cap \Hc(f_u)\}$.
Unfortunately, such maximum delay is neither deterministic nor known a priori. 

We propose a scheme where each user $u$ estimates its local delays by monitoring 
its delivery times in a sliding window spanning a fixed number of time slots. 
In addition, users can also skip a chunk if, by doing so, a sufficiently large jump-up in the number of ordered chunks 
in the playback buffer is achieved.  For instance, in Table \ref{chunkarriv-tab} and Fig.~ \ref{chunkarriv-fig}, the chunk which 
comes $4^{\mathrm{th}}$ in the ordered  sequence arrives at the end of time slot $11$. 
However, chunks numbered $5, 6, 7$ and $8$ arrive before slot $11$ but cannot be played 
since $4$ is missing. More generally, if chunk $4$ were to arrive with a delay such that the number of chunks which arrive before $4$ but 
come later in the ordered sequence becomes large, then the user could either continue waiting for the missing chunk 
and incur a stall event,  or skip chunk $4$ from playback and take advantage of the many already received chunks. 

\begin{table*}
\caption{Arrival times of chunks}
\centering
\begin{tabular}{|c | c | c | c | c | c | c | c | c | c | c | c | c | c|}
\hline
Chunk number & 1 & 2 &  3 &  4 &  5 & 6 & 7 & 8 & 9 & 10 & 11 & 12 & 13\\
\hline
Arrival time & 3 & 4 & 5 & 11 & 6 & 8 & 9 & 10 & 12 & 13 & 16 & 15 & 14  \\
\hline
\end{tabular}
\label{chunkarriv-tab}
\end{table*}

Let $t_k$ denote the time slot in which a user requests the $k^{\mathrm{th}}$ chunk and let $A_k$ be the time slot 
in which the chunk arrives at the user playback buffer. The delay for chunk $k$ is $W_k = A_k - t_k$. 
Without loss of generality, consider a user $u$ starting its streaming session at time $t = 1$. In the proposed scheduling policy, 
user $u$ requests one chunk per scheduling time, sequentially and possibly from different helpers, such that $t_k = k$. 
Since the chunks are downloaded from different helpers with different queue lengths, 
they may be received out of order. For example, it may happen that $A_k < A_j$ for some $j < k$. 
Hence, chunk $k$ cannot be played until all chunks $j$ for $j < k$ are also received. 
We say that a chunk $k$ becomes {\it playable} when all the chunks $j \leq k$ are received. 
Let $P_k$ denote the time when chunk $k$ becomes {\it playable}. Then, we have:
\begin{align} 
P_k = \max \{ A_1, A_2, \ldots, A_k\}.
\end{align} 
The proposed policy consists of two parts: skipping chunks from playback 
and buffering policy.  We examine these two features separately in the following.

\subsection{Skipping chunks from playback}

Prior to slot $t$, the set of playable chunks is $\{k: P_k \leq t-1\}$ and 
\begin{align}
k^*_{t-1} = \max \{k: P_k \leq t-1\}
\label{last-playable}
 \end{align} 
is the highest-order chunk in the ordered sequence of playable chunks. 
At the end of slot $t$, user $u$ considers the set $\Cc_t$ of all chunks  
which have arrived before or during slot $t$  
and which come later than $k^*_{t-1}$ in the ordered sequence of playback.  
The set $\Cc_t$ is given by:
\begin{align} 
\Cc_t = \{k: k>k^*_{t-1}, \; A_k \leq t \}. 
\label{available}
\end{align}
The next available chunk with order larger than $k^*_{t-1}$ is given by: 
\begin{align}
k^-_t = \min \{k: k > k^*_{t-1}, \; A_k \leq t\}.
\label{first-available}
\end{align}
Let $\Cc^*_t \subseteq \Cc_t$ be the set of chunks which become playable at the end of slot $t$, i.e., 
\begin{align}
\Cc^*_t =  \{k: A_k \leq t, P_k = t\}.
\end{align}
If $k^-_t$ comes next to $k^*_{t-1}$ in the playback order (i.e. if $k^-_t = k^*_{t-1}+1$), then $\Cc^*_t$ is non-empty 
and all the chunks $k \in \Cc^*_t$ can be added to the playback buffer. 
Further, $k^*_t$ is recursively updated as:
\begin{align}
k^*_t = k^*_{t-1} + |\Cc_t^*|.
\end{align}

Denoting the increment in the size of the playback buffer at the end of slot $t$ by $\Lambda_t$, we have in this case that $\Lambda_t = |\Cc_t^*|$. On the other hand, if $k^-_t$ is not the immediate successor of $k^*_{t-1}$ in the playback order (i.e. if $k^-_t > k^*_{t-1} + 1$), 
then there is no chunk in $\Cc_t$ which becomes playable at the end of slot $t$ and therefore $\Cc^*_t = \emptyset$.  
In this case, the algorithm compares $|\Cc_t|$ with a threshold $\rho$ in order to decide whether 
it should wait further for the missing chunk $k^*_{t-1} + 1$ or skip it in order to increase the playback buffer anyway.  
The intuition behind such a decision is that it is worthwhile to skip a chunk 
if skipping such a chunk results in a large jump in the playback buffer size. The size of this possible jump can be exactly computed from $\Cc_t$ as follows:  assuming
$k^-_t = k^*_{t-1} + 2$, if we skip chunk $k^*_{t-1}+1$, then the increase in the playback buffer is given by the size of the set:
\[  \{j \geq 2 \; : \; k^*_{t-1}+ i \in \Cc_t~\forall~2 \leq i \leq j \}. \]
We therefore propose the following policy: if $|\Cc_t| \leq \rho$ (where $\rho$ is a parameter $> 0$), then wait for chunk $k^*_{t-1}+1$ and let $k^*_t = k^*_{t-1}$. 
Otherwise, if $|\Cc_t| > \rho$, the increase of the playback buffer is worthwhile and therefore it is useful to
skip chunk $k^*_{t-1} + 1$.  In this case, if $k^-_t = k^*_{t-1} + 2$, then  $k^*_t$ is updated as
\begin{align}
k^*_{t} = k^*_{t-1} + \max \{j: k^*_{t-1}+i~\forall~2 \leq i \leq j \}
\end{align}
and all the chunks numbered from $k^*_{t-1}+2$ to $k^*_t$ are made playable at the end of slot $t$ and added to the playback buffer. We therefore have $\Lambda_t = |\{j>2: k^*_{t-1}+i~\forall~2 \leq i \leq j \}|$ in this case.
Instead, if $k^-_t > k^*_{t-1}+2$, then the user skips chunk $k^*_{t-1}+1$ and starts waiting for chunk $k^*_{t-1}+2$. Only a single chunk is allowed to be skipped per slot because skipping multiple chunks might cause damage to the quality of experience of the user. Note that in this case,   $k^*_t$ is updated to $k^*_{t -1}+1$ even though the chunk $k^*_{t-1}+1$ is missing and is not playable.  This is to ensure that when chunk $k^*_{t-1}+2$ is received, it is considered playable despite the fact that chunk $k^*_{t-1}+1$ is missing. Also note that  there is no increment in the playback buffer (i.e., $\Lambda_t = 0$) in this case because there is no new chunk which becomes playable. Note that choosing $\rho = \infty$ corresponds to the case when no chunk is skipped.

\subsection{Pre-buffering and re-buffering}

The goal here is to determine the delay $T_u$ after which user $u$ should start playback, with respect to the time at which 
the first chunk is requested (beginning of the streaming session).  Intuitively, choosing a large $T_u$ makes the  
buffer underrun rate small. 
However, a too large $T_u$ is very annoying for the user's quality of experience.  
From the chunk skipping strategy seen above, we know that $\Lambda_t$ is the number of new chunks added to the playback buffer 
at the end of slot $t$. We define the size of the playback buffer $\Psi_t$ as the number of playable chunks in the buffer not yet played. 
Without loss of generality,  assume again that the streaming session starts at $t = 1$. 
Then, $\Psi_t$ is recursively given by the updating equation:
\begin{align}
\Psi_t = \max \left \{ \Psi_{t-1}  -  1\{t > T_u\}, 0 \right \} + \Lambda_t.
\end{align}
From the qualitative discussion on the evolution of the playback buffer at the beginning of Section
\ref{sec:prebuffering}, we notice that the longest period during which $\Psi_t$ is not incremented (in the absence of chunk skipping decisions)
is given by the maximum delay $W_k$ to deliver chunks.  In addition, we note that each user $u$ needs to adaptively estimate $W_k$ in order to choose $T_u$. In the proposed method, user $u$ calculates for every chunk $k$ the corresponding delay
$W_k = A_k - t_k$. Notice that the delay of chunk $k$,  can be calculated only at time $A_k$, i.e., 
when the chunk is actually delivered. At each time $t = 1,2,\ldots$, user $u$ calculates the maximum observed delay $E_t$  in a sliding 
window of size $\Delta$, (in all the numerical experiments in the sequel, we use $\Delta =10$) by letting:
\begin{align}
E_t = \max \{ W_k \; :  ~t-\Delta+1 \leq A_k \leq t \}.
\label{del-window}
\end{align}
Finally, user $u$ starts its playback when $\Psi_t$ crosses the level $\xi E_t$, i.e.,  
\begin{align}
T_u = \min \{ t : ~\Psi_t \geq \xi E_t \}.
\end{align}
If we have $\Psi_t = 0$ for some $t > T_u$, a stall events occurs and the algorithm enters a re-buffering phase 
in which the same algorithm presented above  is employed again to determine the new instant $t + T_u + 1$ at which playback is restarted. 
Notice that, with some abuse of notation, we have denoted the re-buffering delay again by $T_u$ although this is re-estimated 
using the sliding window  method at each new stall event. In fact, when a stall event occurs, it is likely that some change in the 
network state has occurred, such that the maximum delay must be re-estimated.

\section{Numerical Experiments, Discussion and Conclusions} \label{sec:simul-dilip}

 
{ In this section, we present two targeted numerical experiments illustrating the particular features of the proposed scheme. 
The first experiment considers the performance under a ``macro-diversity" physical layer, for which the rate scheduling sub-problem
takes on the form (\ref{sucasuca}). We consider a large network with many stationary users and one mobile user moving across the network at constant speed. 
Users alternate between {\it idle} and {\it active} phases of video streaming. Each streaming session (when moving form idle to active state) is initialized
using the pre-buffering scheme described in Section \ref{sec:prebuffering}.  
This simulation demonstrates the dynamic and adaptive nature of the policy in response to 
VBR video coding  and users joining or leaving the system at arbitrary times.
Furthermore, the statistics relative to the streaming session of the mobile user shed light on the ability of the proposed algorithm
to seamlessly discover new helper nodes as the user changes its position across the network. 
The second experiment considers a smaller network formed by four helpers and several users, in a situation of congestion for which most users
are close to one helper. We consider the proposed scheme both under a ``macro-diversity" and under ``unique association" physical layer 
(where in the latter case, the rate scheduling problem takes on the form (\ref{mwsr-matching})) and compare its performance
with a naive approach with max-SINR user-helper association, representative of today's baseline technology. 
}

As described in Section \ref{sec:intro-dilip}, the helpers could be base stations 
connected to some video server through a wired backbone, or they could be dedicated wireless nodes with local caching capacity. 
For the sake of simplicity and replicability of our numerical results, here we assume that  each helper has available the whole video library. 
Therefore, for any request $f_u$ we have $\Nc(u) \cap \Hc(f_u) = \Nc(u)$. 
We use the utility function $\phi_u(x) = \log(x)$ for all $u \in \Uc$ (i.e., we use $\alpha$-fairness with $\alpha = 1$ \cite{mo2000fair}).  
As described in Section~\ref{sec:sysmodel-dilip}, a scheduling slot duration of $0.5$s
and a total available system bandwidth of $W = 18$ MHz yield $10^5$ LTE resource blocks per slot \cite{sesia-LTE}.
The total number of channel symbols $n$ in a scheduling slot is $10^5 \times 84$.
We assume that each user $u$ has an edge to every helper $h$ which satisfies 
$n C_{hu}(t) > 1$ Mb (i.e., at least 2 Mb/s of peak rate).
 
The path loss coefficients $g_{hu}(t)$ between helper $h$ and user $u$ are based on the 
WINNER II channel model \cite{winner4}.  In particular, we let
\[ g_{hu}(t) = 10^{-\frac{\mathrm{PL}(d_{hu}(t))}{10}}, \]
where $d_{hu}(t)$ is the distance from helper $h$ to user $u$ at time $t$, and where
\begin{align}
\mathrm{PL}(d) = A\log(d) + B + C\log(f_0/5) + \chi_{\mathrm{dB}}. \label{winner}
\end{align}
In (\ref{winner}),  $d$ is expressed in meters, 
the carrier frequency $f_o$ in GHz,  and $\chi_{\mathrm{dB}}$ denotes a shadowing log-normal variable with variance $\sigma_{\mathrm{dB}}^2$. 
The parameters $A,B,C$ and $\sigma_{\mathrm{dB}}^2$ are scenario-dependent constants. 
Among the several models specified in WINNER II  we chose the A1 model in \cite{winner4}, representative of a small-cell scenario.  
In this case, $3 \leq d \leq 100$, and the model parameters are given by $A=18.7$, $B=46.8$, $C=20$, $\sigma_{\mathrm{dB}}^2 = 9$
in line-of-sight (LOS) condition, or $A=36.8$, $B=43.8$, $C=20$, $\sigma_{\mathrm{dB}}^2 = 16$ in non-line-of-sight (NLOS) condition. 
For distances less than 3 m, we extended the model by setting $\mathrm{PL}(d) = \mathrm{PL}(3)$.
Each link is in LOS or NLOS independently and at random, with probability $p_l(d)$ and $1 - p_l(d)$, respectively, where
\begin{equation*}
p_l(d) = \left \{ \begin{array}{ll}
1 & d\leq 2.5\mathrm{m} \\
 1-0.9(1-(1.24-0.6\log(d))^3)^{1/3} & \mathrm{otherwise} \end{array} \right .
\end{equation*} 
Every helper transmits at fixed power level $P=10^8$. 

Using Jensen's inequality in (\ref{rateconst2}) to replacing the denominator of the SINR 
term with its average (this will be the average received inter-cell interference power), 
and the  fact that the small-scale fading coefficients $s_{hu}$ are $\sim \Cc\Nc(0,1)$,
the peak achievable rates can be lower-bounded by the closed-form expression
$C_{hu}(t) = e^{1/\Gamma_{hu}(t)} {\rm Ei} \left (1 , \frac{1}{\Gamma_{hu}(t)} \right )$, 
where ${\rm Ei}(1,x) = \int_x^\infty \frac{e^{-t}}{t} dt$ for $x \geq 0$, 
and $\Gamma_{hu}(t) = \frac{P_hg_{hu}(t)}{1+\sum_{h^{'}\neq h}   P_{h^{'}}g_{h^{'}u}(t)}$. 
This formula, which provides a very accurate lower bound to (\ref{rateconst2}) when the SINR denominator in (\ref{rateconst2}) contains many 
independent terms, is an {\em achievable rate}\footnote{A lower bound to an achievable rate is obviously achievable.} 
and is used in the numerical results of this section. 

\begin{figure}[h]
\subfloat[Bitrate profile]{\includegraphics[scale=0.45]{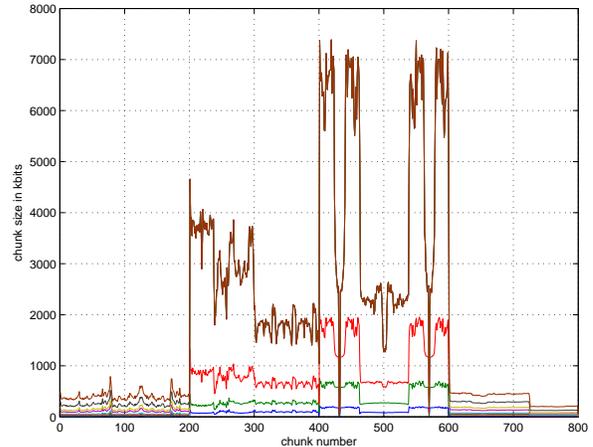}
\label{bitperchunk}} \\
\subfloat[Quality profile]{
\includegraphics[scale=0.45]{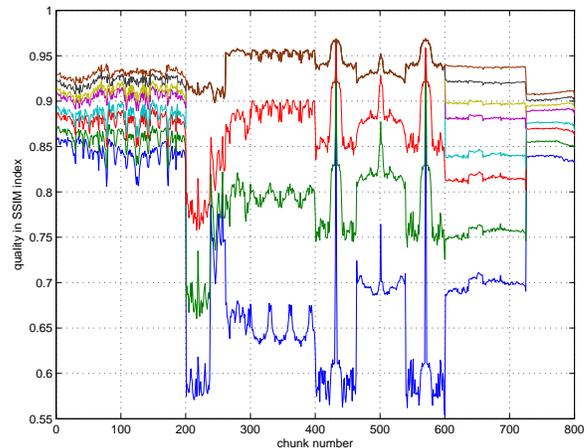}\label{ssimperchunk} 
}
\caption{Rate-quality profile of the test video sequence used in our simulations.}
\end{figure}

We assume that all the users request chunks successively from VBR-encoded video sequences.  
Each video file is a long sequence of chunks, each of duration $0.5$ seconds 
and with a frame rate $\eta = 30$ frames per second. 
We consider a specific video sequence formed by $800$ chunks, constructed using $4$ video clips 
from the database in~\cite{video-samples}, each of length $200$ chunks. 
The chunks  are encoded into different quality modes. 
Here, the quality index is measured using the {\em Structural SIMilarity} (SSIM) index defined in \cite{ssim}.
Fig.s~\ref{bitperchunk} and \ref{ssimperchunk} show the size in kbits and the SSIM values as a function of the chunk index, respectively, 
for the different quality modes. In our experiments, the chunks from $1$ to $200$ and $601$ to $800$ are encoded into $8$ quality modes, 
while the chunks numbered from $201$ to $600$ are encoded in $4$ quality modes. 
In both the experiments in the sequel, each user starts its streaming session of $1000$ chunks from some arbitrary position 
in this reference  video sequence and successively requests $1000$ chunks by cycling through the sequence. 

\subsection{Experiment 1}

\begin{figure}[h]
\centering
\includegraphics[scale=0.45]{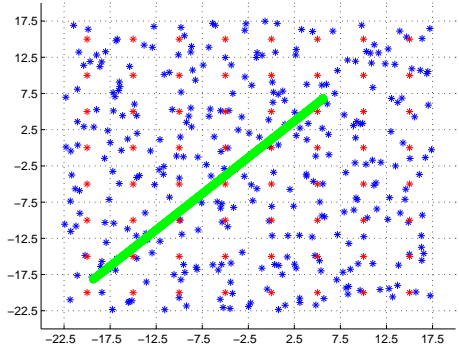}
\caption{Toplogy (the green line indicates the trajectory of the mobile user in Experiment 1).}
\label{topology}
\end{figure}

In the large network experiment, we consider a $40$m $\times 40$m square area divided into $8\times8$ small square cells of side length $5$m 
as shown in Fig.~\ref{topology}.  A helper is located at the center of each small square cell. 
The network includes $319$ randomly placed stationary users 
and one mobile user whose trajectory is indicated by the green line. 
At $t=0$, the mobile user starts a video streaming session of $1000$ chunks. 
Simultaneously, it starts moving along the trajectory and stops after it requests the $1000^{\mathrm{th}}$ chunk. 
It doesn't request any more chunks after it  stops moving. As the user moves through its trajectory, the new path loss coefficients $g_{hu}(t)$ are 
calculated using the Winner II model said above,  leading time-varying peak link rates $C_{hu}(t)$. 
The remaining $319$ users in the system are stationary throughout the simulation period and alternate between {\it idle} and {\it active} phases 
of video streaming.  At $t=0$, all the stationary users are idle and each one of them independently starts a streaming session with 
probability  $p=0.005$ at every slot.  Thus, the time for which a user stays idle is geometrically distributed with mean $\frac{1}{p} = 200$ slots.
Once a user starts a streaming session, it stays {\it active} during $1000$ video chunks.
After finishing the requests, it goes back into the {\it idle} state and may start a new session after an independent and random geometrically distributed 
idle time. We simulate the { proposed scheme under the macro-diversity physical layer} 
for $3000$ slots for fixed values of the key parameters $V$, $\xi$ and $\rho$ set to $10^{13}, 25$ and $50$ respectively.  
These values have been chosen after extensive simulation and yields a good behavior of the scheduling policy. 
In general, the policy parameters have to be tuned to the specific network environment.  

\begin{figure*}
\centering
\subfloat[]{
\includegraphics[width = 70 mm, height = 46mm]{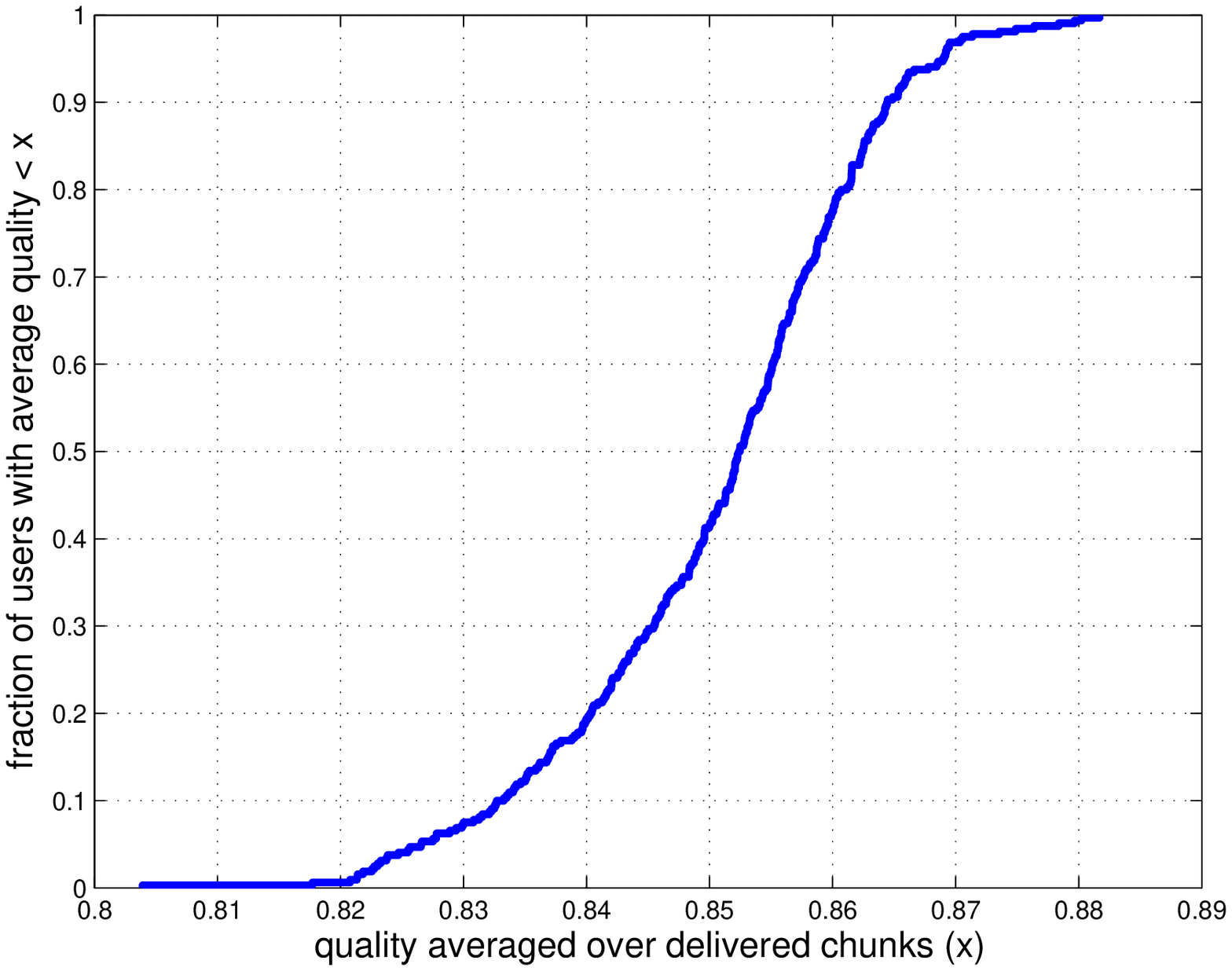}
\label{ssim-cdf-v7}
}
\subfloat[]{
\includegraphics[width = 70 mm, height = 46mm]{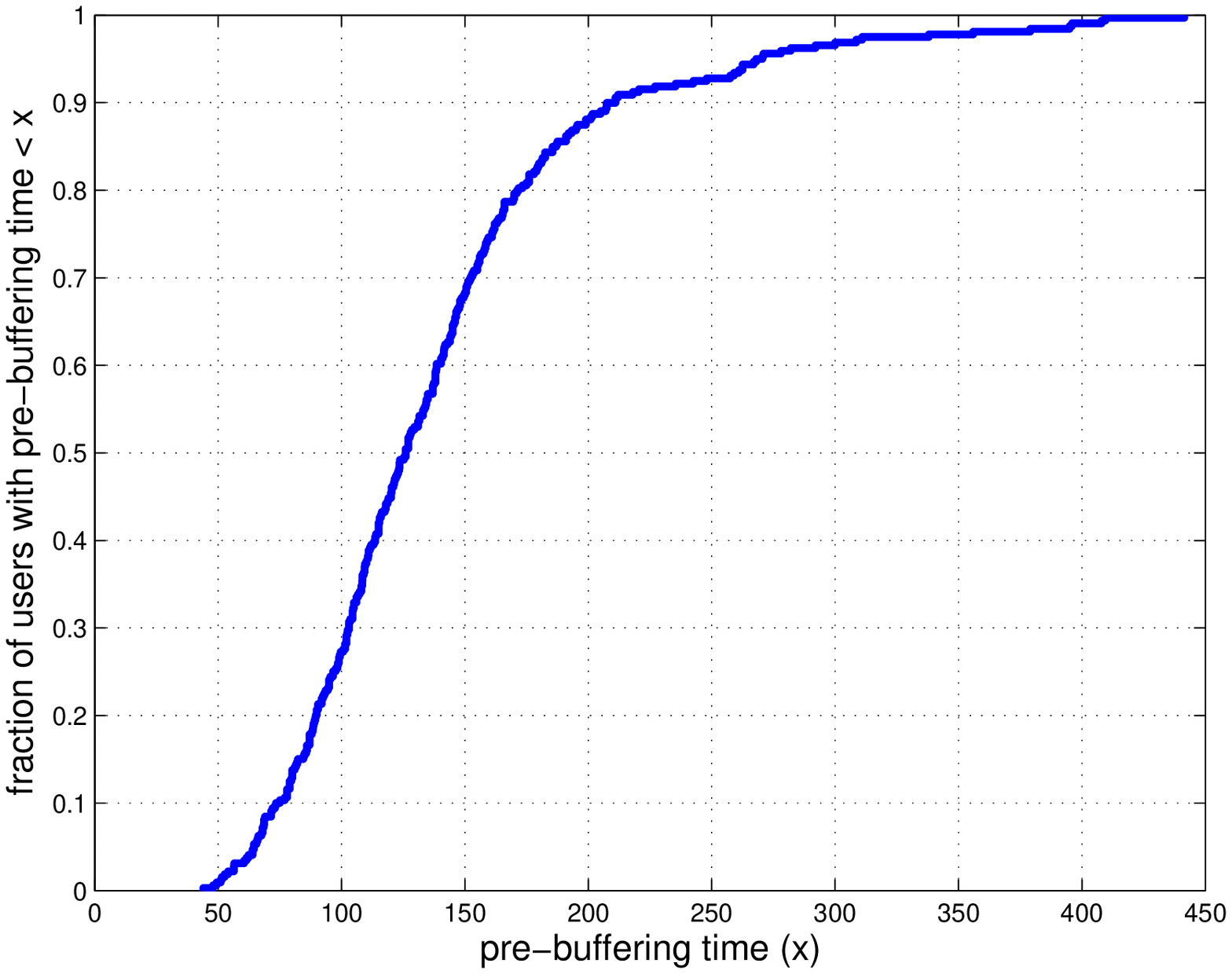}
\label{prebuff-cdf-v7}
}\\
\subfloat[]{
\includegraphics[width = 70 mm, height = 46mm]{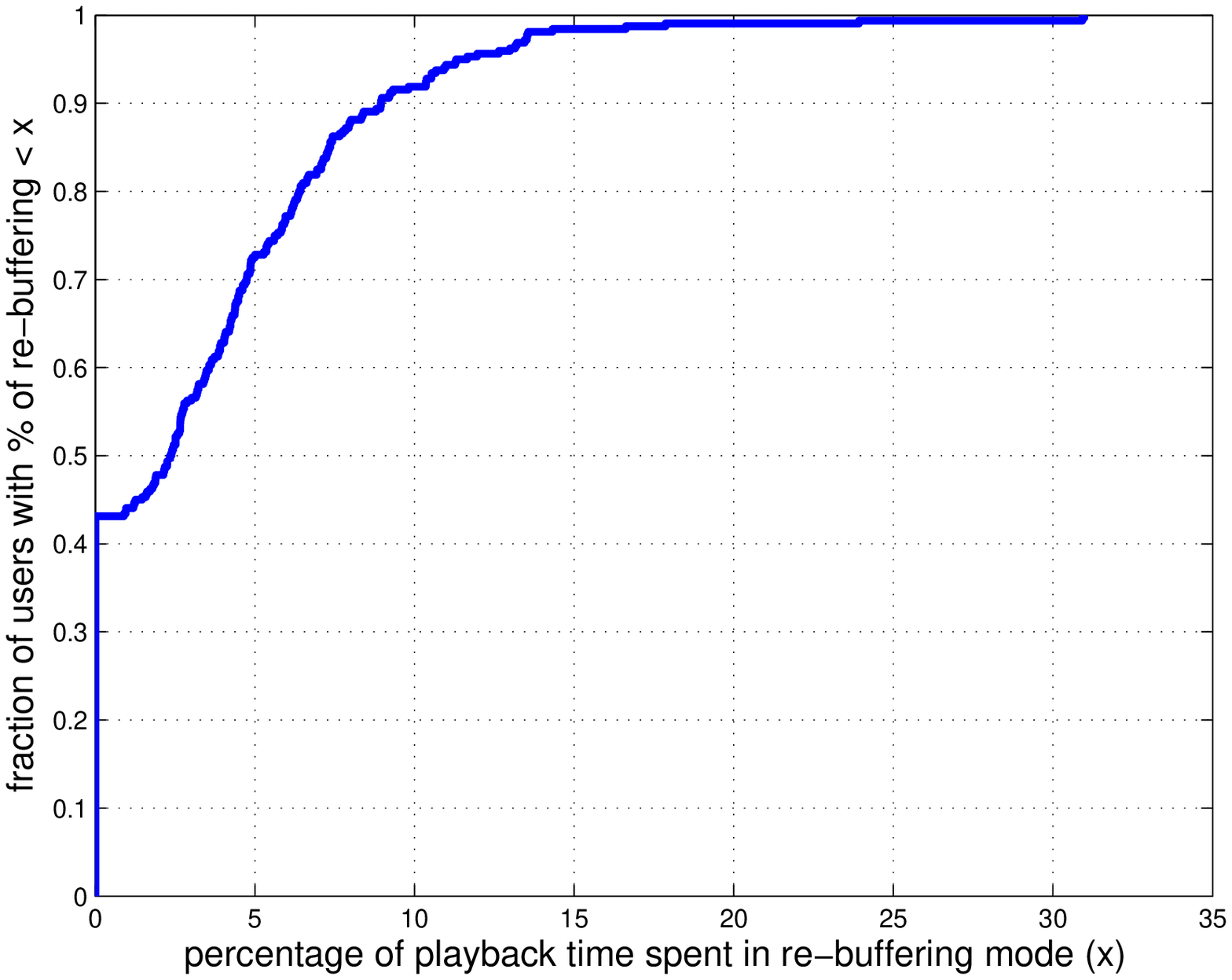}
\label{rebuff-cdf-v7}
}
\subfloat[]{
\includegraphics[width = 70 mm, height = 46mm]{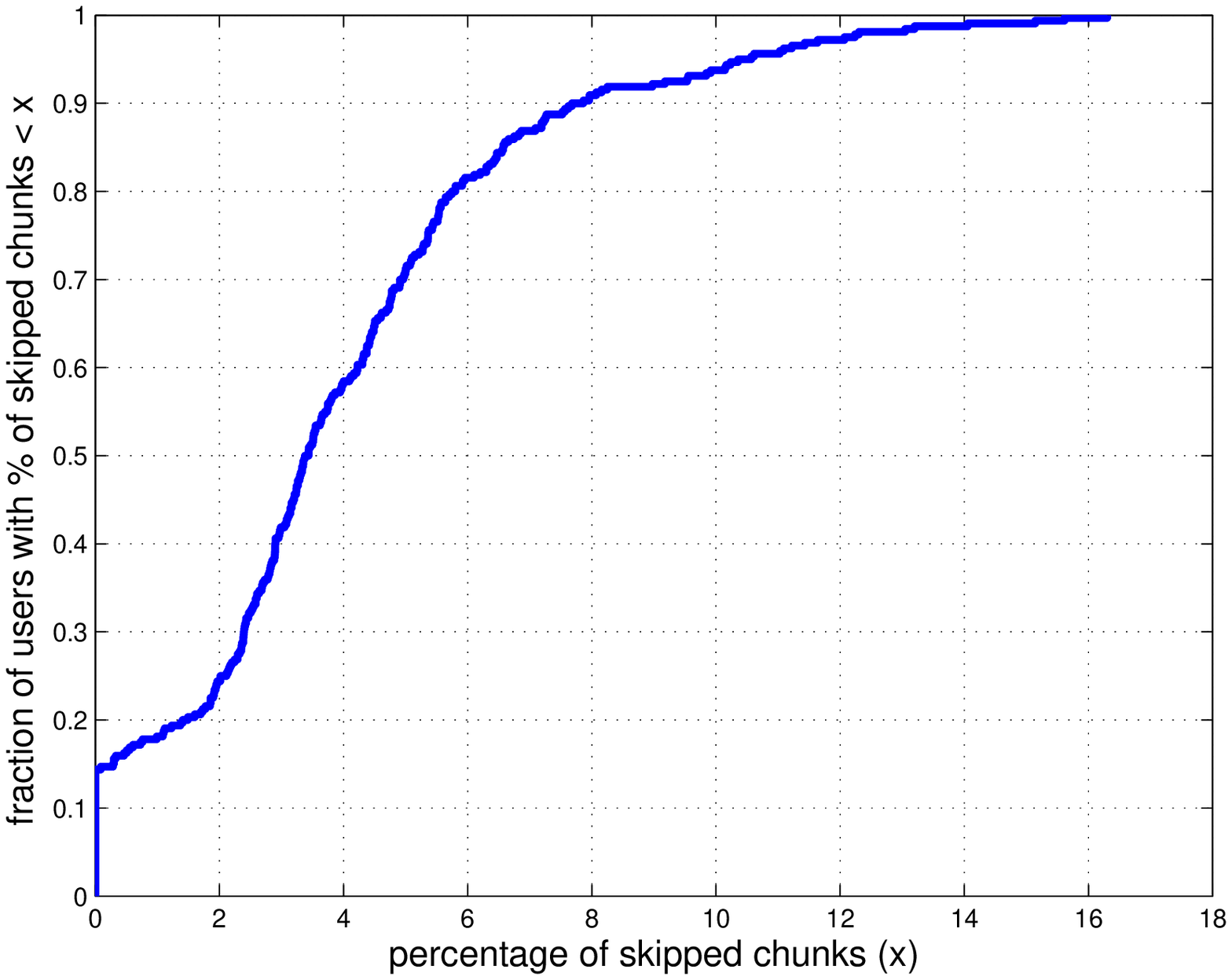}
\label{skip-cdf-v7}
}\\
\subfloat[]{
\includegraphics[width = 70 mm, height = 46mm]{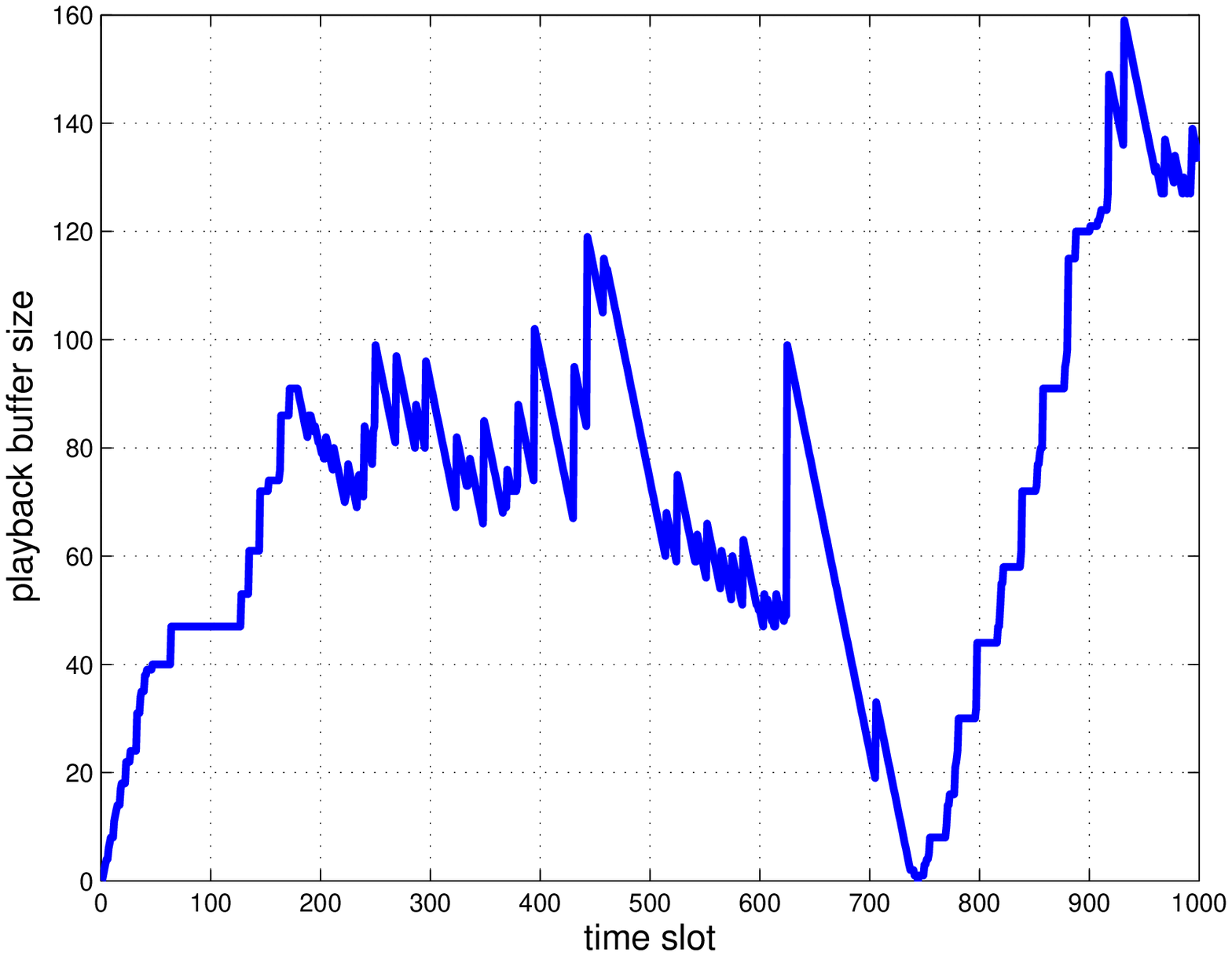}
\label{mobile-buffer-v7}
}
\subfloat[]{
\includegraphics[width = 70 mm, height = 46mm]{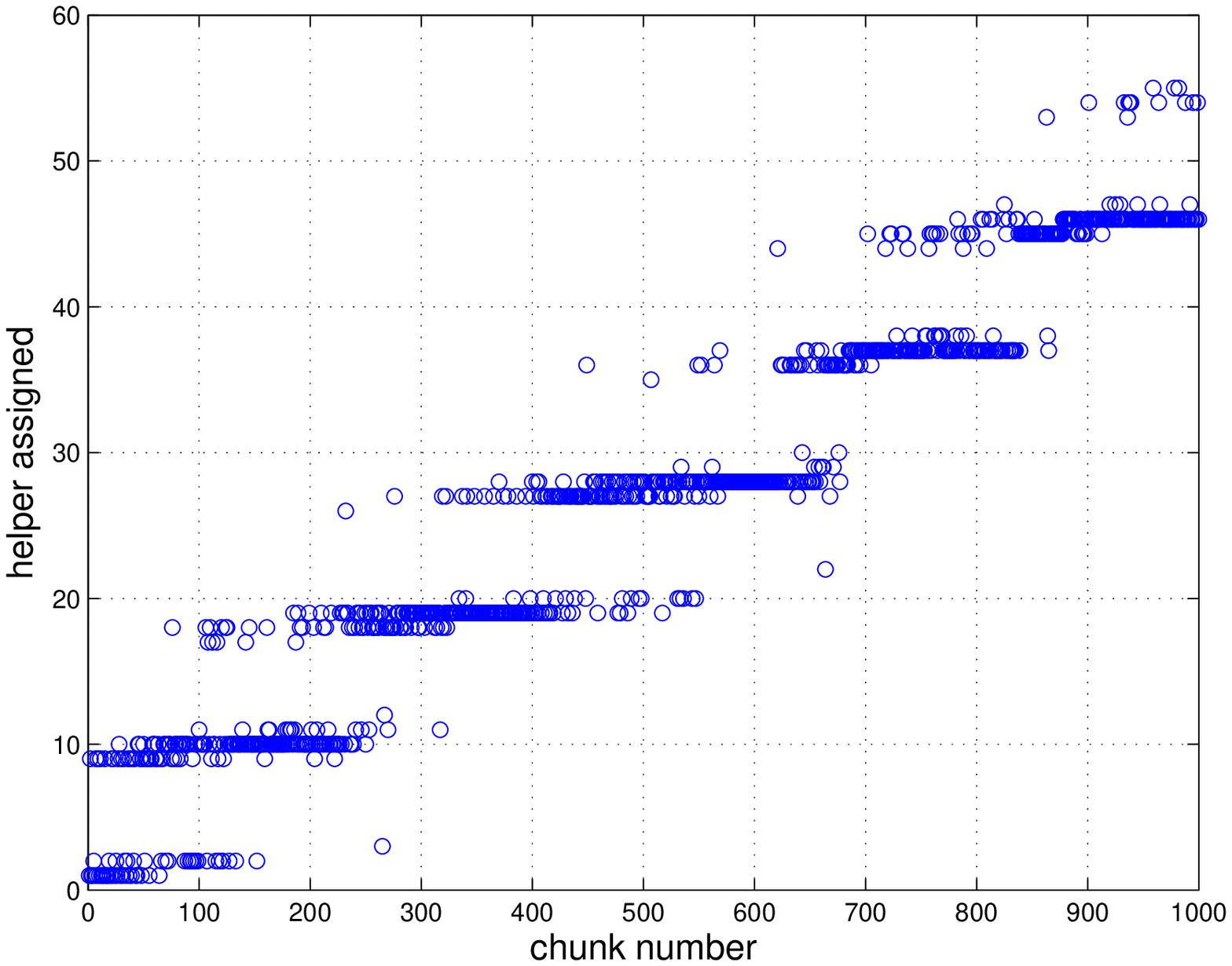}
\label{seamless-download-v7}
}
\caption{CDFs of different performance metrics for Experiment 1.}
\end{figure*}

We show the results in terms of the empirical CDF (over the user population) of the following metrics:
 1) The percentage of skipped chunks spanning multiple streaming sessions of each user (Fig.~\ref{skip-cdf-v7});
 2) The quality (SSIM) averaged over the delivered chunks spanning multiple streaming sessions of each user (Fig.~\ref{ssim-cdf-v7});
  3) The initial pre-buffering time (in number of slots) is calculated for each streaming session (Fig.~\ref{prebuff-cdf-v7});  
  4) The percentage of  time spent in {\it re-buffering} mode is calculated with respect to the total playback time spanning multiple streaming 
  sessions of each user (Fig.~\ref{rebuff-cdf-v7}).

Focusing on the mobile user, we observe that the percentage of skipped chunks is $16 \%$ and the pre-buffering 
time is $180$ time slots (i.e., 1min). 
Fig.~\ref{mobile-buffer-v7} shows the evolution of the playback buffer $\Phi_t$ over time. We notice that there is only one interruption (stall event)
in the entire streaming session. The quality (SSIM) averaged over the delivered chunks is observed to be a high 
value of $0.87$ (the maximum being 1.0).  The helpers are numbered from $1$ to $64$, left to right and bottom to top, in Fig.~\ref{topology}. 
In Fig.~\ref{seamless-download-v7}, we plot the helper index  providing chunk $k = 1, \ldots, 1000$ vs. the chunk index. We can observe that as the user moves slowly along the path, the policy ``discovers'' adaptively the current neighboring helpers and downloads chunks from them in a seamless fashion.
Overall, these results demonstrate the dynamic and adaptive nature of the proposed policy in response to user mobility, variable bit-rate video coding, 
and users joining or leaving the system at arbitrary times.

\subsection{Experiment 2}

\begin{figure*}[t]
\subfloat[]{
\includegraphics[scale=0.4]{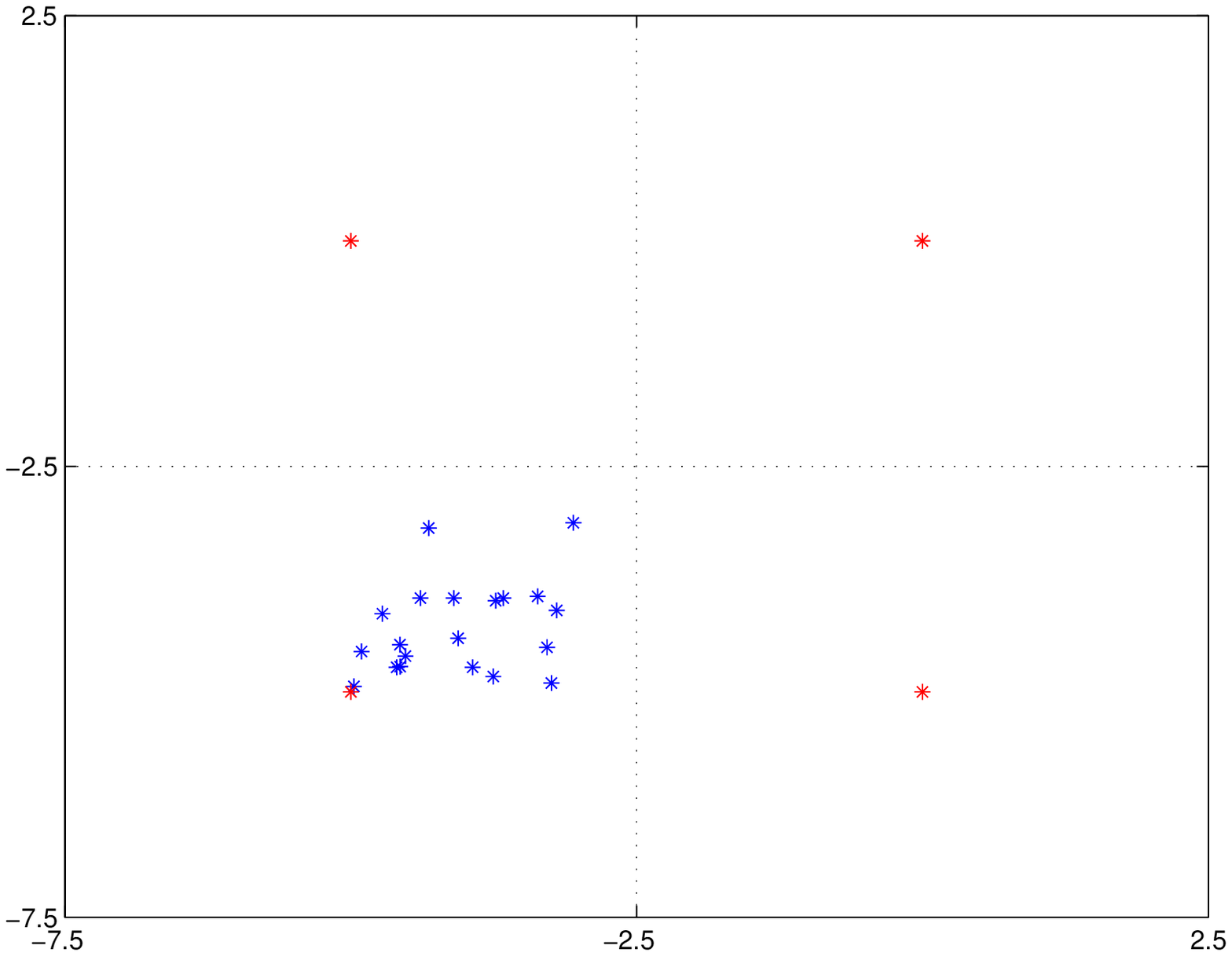}
\label{topology-exp2}
}
\subfloat[]{
\includegraphics[scale=0.4]{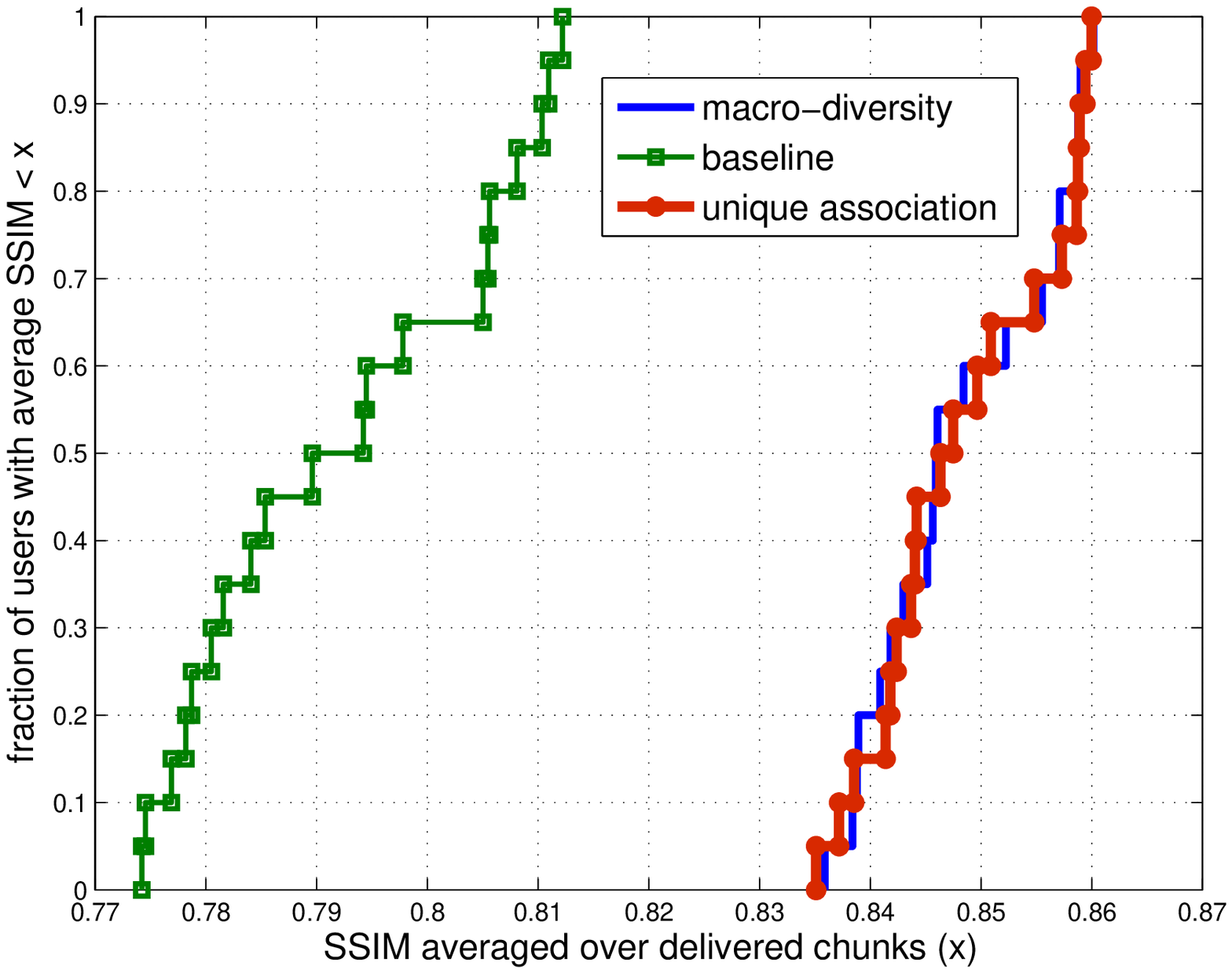}
\label{ssim-cdf-v7-exp2}
}\\
\centering
\subfloat[]{
\includegraphics[scale=0.4]{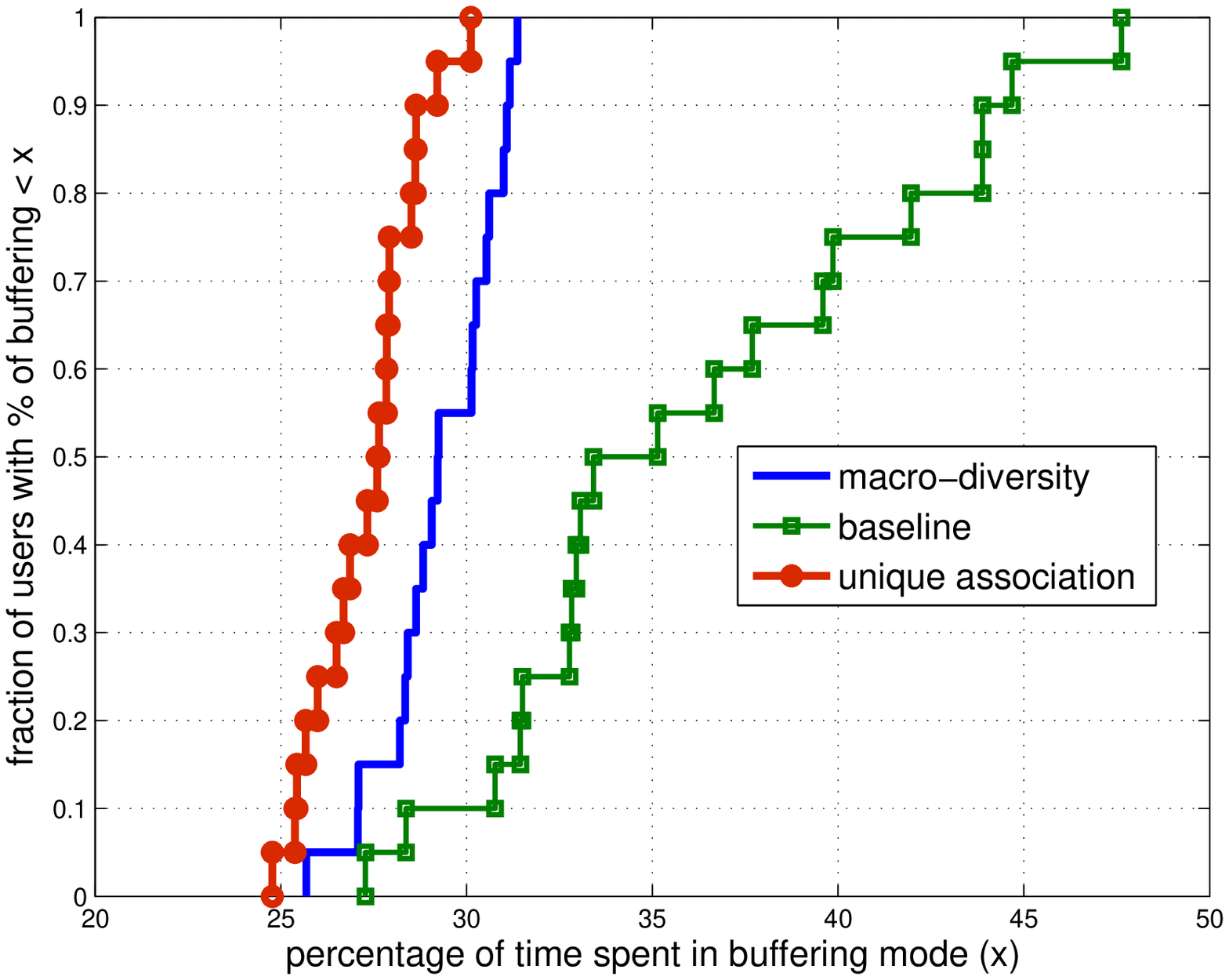}
\label{cdf-prebuff-v7-exp2}
}
\caption{Topology and CDFs of different performance metrics for Experiment 2.}
\end{figure*}

In this experiment, we focus on a smaller network with $4$ helpers and $20$ stationary users as indicated in Fig.~\ref{topology-exp2}. 
The dimensions used for the topology are the same as in Fig.~\ref{topology} where each of the $4$ helpers is located at the centre of a $5$m $\times$ $5$m 
square cell and the overall area of the system is $10$m $\times$ $10$m. We consider a situation where the $20$ users in the system are located 
close to the same helper, as indicated in Fig.~\ref{topology-exp2}. We choose this non-uniform user distribution in order to investigate the 
load balancing property of the proposed policy in contrast to a naive scheme that allocates users to helpers based on maximum 
signal strength (or, equivalently, based on maximum SINR). 
In this experiment, all the $20$ users start their streaming session simultaneously at $t=0$, and stop after  $1000$ requested chunks. 
A baseline scheme, representative of current WLAN technology, performs client-based user-helper association, i.e., 
every user $u$ chooses helper $h^*_u(t) = \mbox{argmax}\left \{ C_{hu}(t): h \in \Nc(u) \right \}$. Then, the streaming process takes place 
accordingly by adapting  the requested video quality according to DASH \cite{sanchez2011improved,sanchez2011idash}. We have emulated this situation by 
applying the same video quality level decisions as in (\ref{quality-level-decision}), with user-helper association as given above. 
{ 

We provide results for the proposed schemes with both ``macro-diversity'' and ``unique association''. 
In order to simulate the unique association scheme, we solve the LP relaxation of (\ref{mwsr-matching}) in every slot using the standard linear 
programming solver of MATLAB. In practice, this can be implemented by a centralized network controller.
The results are shown in the form of empirical CDF (over the user population) of: 1) SSIM averaged over the chunks (Fig.~\ref{ssim-cdf-v7-exp2}); 
2) fraction of slots spent in buffering mode (including pre-buffering and re-buffering periods) (Fig.~\ref{cdf-prebuff-v7-exp2});  
We notice that the proposed policy, both under macro-diversity and unique association, improves over the baseline scheme in terms of the 
video quality metric  and the fraction of slots spent in buffering mode.  
This is because the baseline scheme a priori fixes the association of a user to the helper with best peak link rate,  while the proposed schemes yield 
better load balancing by allowing each user to dynamically select the best helper in its neighborhood based on the congestion control decision 
(\ref{helper-selection}), which takes into account the length of all queues ``pointing'' at the user itself. 
In addition, we notice that though the macro-diversity and the unique association schemes differ significantly in terms of implementation, the difference
in terms of performance is small. This shows that 1) even in such a small cell scenario, macro-diversity does not provide a large gain over unique association;\footnote{Notice that in a macro-cell scenario, where most users are in good SINR conditions to at most one base station, macro-diversity would yield an even smaller performance gain over  unique association.}  
2) the major source of gain of the proposed scheme over the base line scheme is due to its seamless load balancing property;
3) the main advantage of a macro-diversity physical layer over a physical layer where unique association is enforced consists of 
the simplicity of the decentralized nature of rate scheduling subproblem (\ref{sucasuca}) over the
centralized maximum weighted matching solution (\ref{mwsr-matching}). 
}

%

\section*{Acknowledgment}
The authors would like to thank Hilmi Enes Egilmez and Prof. Antonio Ortega for providing a scalable encoded variable bitrate video sequence for the experiments and also for several useful discussions.
 The authors would also like to thank the anonymous reviewers whose comments greatly helped to clarify certain technicalities related to queue delays and the role of pre-buffering, which were not precisely explained in the early versions of the paper.

\appendices

\section{Proof of Lemma \ref{equivalence}}  \label{proof-lem}

Let $\phi^{\mathrm{opt}}_1$ and $\phi^{\mathrm{opt}}_2$ be the optimal solutions of problems (\ref{NUMproblem}) and 
(\ref{maxutiltrans}) -- (\ref{feasibleoptionstrans}), respectively. 
Formally, $\phi^{\mathrm{opt}}_1$ is the supremum objective function value over all algorithms that satisfy the constraints of problem (\ref{NUMproblem}).  
The value $\phi^{\mathrm{opt}}_2$ is defined similarly.    
Now, fix $\epsilon > 0$ and let $a^*(t)$ be a policy that satisfies all constraints of the transformed problem 
(\ref{maxutiltrans}) -- (\ref{feasibleoptionstrans}) and achieves a utility not smaller than $\phi^{\mathrm{opt}}_2 - \epsilon$.
We have
\begin{equation}
\phi^{\mathrm{opt}}_2 - \epsilon \leq \sum_{u \in \Uc}\overline{\phi_u(\gamma_u^*)} 
\stackrel{(a)}{\leq} \sum_{u \in \Uc}\phi_u(\overline{\gamma_u^*}) 
\stackrel{(b)}{\leq} \sum_{u \in \Uc}\phi_u(\overline{D_u^*}) 
\stackrel{(c)}{\leq} \phi_{\mathrm{opt}}^1, \label{forward}
\end{equation}
where (a) follows from Jensen's inequality applied to the concave function $\phi_u(\cdot)$, 
(b) follows by noticing that the policy $a^*(t)$ satisfies the constraint (\ref{gammaconst}) and 
$\phi_u(\cdot)$ is non-decreasing, and (c) follows from the fact that since $a^*(t)$ is feasible for 
problem (\ref{maxutiltrans}) -- (\ref{feasibleoptionstrans}),  then it also satisfies the constraints of problem (\ref{NUMproblem})
and therefore it is feasible for the latter.  As this holds for all $\epsilon > 0$, we conclude that $\phi^{\mathrm{opt}}_2 \leq \phi^{\mathrm{opt}}_1$. 

Now, let $a^{'}(t)$ be a policy for the original problem (\ref{NUMproblem}),  achieving a utility not smaller than $\phi_1^{\mathrm{opt}} - \epsilon$. 
Since $a^{'}(t)$ is feasible for (\ref{NUMproblem}), it also satisfies the constraints 
(\ref{qstableconsttrans}), (\ref{feasibleoptionstrans}) of the transformed problem.  
Further, we choose $\gammav^{'}(t) = \overline{\Dm^{'}}$ for all time $t$. Such choice of $\gammav^{'}(t)$ together 
with the policy $a^{'}(t)$ forms a feasible policy for problem (\ref{maxutiltrans}) -- (\ref{feasibleoptionstrans}). Therefore:
\begin{align}
\phi^{\mathrm{opt}}_1 - \epsilon \leq \sum_{u \in \Uc}\phi_u(\overline{D_u^{'}}) = \sum_{u \in \Uc}\overline{\phi_u(\gamma_u^{'})} \leq \phi^{\mathrm{opt}}_2. \label{reverse}
\end{align}
As this holds for all $\epsilon > 0$, we conclude that $\phi^{\mathrm{opt}}_1 \leq \phi^{\mathrm{opt}}_2$.
Thus, (\ref{forward}) and (\ref{reverse}) imply that  $\phi^{\mathrm{opt}}_1 = \phi^{\mathrm{opt}}_2$ and, by comparing the constraint, 
it is immediate to conclude that an optimal policy for the transformed problem can be directly turned into an optimal 
policy for the original problem.

\section{Proof of Theorem \ref{main-result} and of Corollary \ref{corcor}}  \label{proof-thm}

As in Section~\ref{derivation}, we consider the following problem, equivalent to (\ref{obj-arbit}) -- (\ref{feas-action-arbit}), 
which involves a sum of time-averages instead of functions of time averages and introduces the auxiliary variables $\gamma_u(t)$:
\begin{align}
\label{constraint} 
\textrm{maximize}  &\;\;\; \frac{1}{T}\sum_{\tau=jT}^{(j+1)T-1} \sum_{u \in \Uc}\phi_u\left(\gamma_u(\tau)\right)   \\
\textrm{subject to}  &\;\;\; \frac{1}{T} \sum_{\tau=jT}^{(j+1)T-1}\left[
kR_{hu}\left(\tau\right)-n\mu_{hu}\left(\tau\right)\right]\leq0 \notag \\
&~~~~~~~~~~~~~~~~~~~~~~~~~~~~~~~~~~~\forall~(h, u) \in \Ec \label{qstable-arbit} \\
& \;\;\;  \frac{1}{T} \sum_{\tau=jT}^{(j+1)T-1}\left[
\gamma_u\left(\tau\right)-D_u\left(\tau\right)\right]\leq0~\forall~ u \in \Uc \label{virt-stable-arbit}\\
& \;\;\; D_u^{\min} \leq \gamma_u(t) \leq D_u^{\max}~\forall~u \in \Uc, \notag \\
&~~~~~~~~~~~~~~~~~~~\forall~t \in \{jT, \ldots, (j+1)T-1\} \label{gamma-limit} \\
& \;\;\; a(t) \in A_{\omegav(t)}~\forall~t~\in~\{jT, \ldots, (j+1)T-1\}. \label{feasactions-trans-arbit}
\end{align}
The update equations for the transmission queues $Q_{hu}~\forall~(h,u) \in \Ec$ and the virtual queues
$\Theta_u~\forall~u \in \Uc$ are given in (\ref{q-update}) and in (\ref{virt-update}), respectively.
 Let ${\bf G}(t)=\left[ {\bf Q}^\transp(t), \Thetav^\transp(t)\right]^\transp$ be the combined queue backlogs column vector, 
and define the quadratic Lyapunov function $L(\Gm(t)) = \frac{1}{2} \Gm^\transp (t) \Gm(t)$.
Fix a particular slot $\tau$ in the $j$-th frame. We first consider the one-slot drift of $L({\bf G}(\tau))$. From~(\ref{drift-bound}), 
we know that
\begin{align}
L({\bf G}(\tau+1))-L({\bf G}(\tau)) \leq  \Kc &+ \left({\bf R}(t)-{\boldsymbol \mu}(\tau)\right)^
\transp{\bf Q}(\tau) \notag \\
&+ \left(\gammav(\tau)-
\Dm(t)\right)^\transp\Thetav(\tau)\label{drift-arbit}
\end{align}
where $\Kc$ is a uniform bound on the term $ \frac{1}{2}\left[{\boldsymbol \mu}^\transp(t){\boldsymbol \mu}(t)+{\bf R}^\transp(t){\bf R}(t)\right]+ \frac{1}{2}\left(\gammav(t)-\Dm(t)\right)^\transp\left(\gammav(t)-\Dm(t)\right)$, that exists under the realistic assumption that the source coding rates, the channel coding rates and the video quality measures are upper bounded by some constants, independent of $t$. We choose $\Kc$ such that
\begin{equation}
\label{B-definition}
 \Kc > 2{\boldsymbol \kappa}^\transp{\boldsymbol \kappa}
\end{equation}
where  $\kappav$ is a vector whose
components are all equal to the same number $\kappa$ and this number is a uniform upper bound on the
maximum possible magnitude of drift in any of the queues (both actual and virtual) in one slot.
With the additional penalty term $-V \sum_{u \in \Uc}\phi_u(\gamma_u(\tau))$ added on both sides of 
(\ref{drift-arbit}), we have the following DPP inequality:
 \begin{align}
 &L({\bf G}(\tau+1))-L({\bf G}(\tau)) -V \sum_{u \in \Uc}\phi_u(\gamma_u(\tau)) \notag \\
 &\leq  \Kc + \left({\bf R}(t)-{\boldsymbol \mu}(\tau)\right)^
\transp{\bf Q}(\tau) + \left(\gammav(\tau)-
\Dm(t)\right)^\transp\Thetav(\tau) \notag \\
&~~~~~~-V \sum_{u \in \Uc}\phi_u(\gamma_u(\tau))\label{driftpluspenalty-arbit}
 \end{align}
Let $\{a(\tau)\}_{\tau=jT}^{(j+1)T-1}$ denote the DPP policy which minimizes the right hand side of the 
{\it drift plus penalty} inequality (\ref{driftpluspenalty-arbit}).
Since it minimizes the expression on the RHS of (\ref{driftpluspenalty-arbit}), any other policy $\{a^*(\tau)\}_{\tau=jT}^{(j+1)T-1}$ comprising of the 
decisions $\{m_u^*(\tau)\}_{\tau=jT}^{(j+1)T-1}$, $\{\Rm^*(\tau)\}_{\tau=jT}^{(j+1)T-1}$, $\{\muv^*(\tau)\}_{\tau=jT}^{(j+1)T-1}$ and $\{\gammav^*(\tau)\}_{\tau=jT}^{(j+1)T-1}$ would give a larger value of the expression. We therefore have
\begin{align}
 &L({\bf G}(\tau+1))-L({\bf G}(\tau))-V \sum_{u \in \Uc}\phi_u(\gamma_u(\tau)) \notag \\
 &\leq \Kc +\left({\bf R}^*(\tau)-\mu^*(\tau)\right)^\transp{\bf Q}(\tau) + \left(\gammav^*(\tau)-
\Dm^*(\tau)\right)^\transp\Thetav(\tau) \notag \\
&~~~~~~-V\sum_{u \in \Uc}\phi_u(\gamma_u^*(\tau)). \label{alpha-star}
\end{align}
Further, we note that the maximum change in the queue length vectors $Q_{hu}(\tau)$ and $\Theta_u(\tau)$ from one slot to the 
next is bounded by $\kappa$. Thus, we have for all $\tau \in \{jT, \ldots, (j+1)T-1\}$
\begin{align}
 |Q_{hu}(\tau)- Q_{hu}(jT)| &\leq (\tau-jT)\kappa~~\forall~(h, u) \in \Ec \label{bounding1}\\
|\Theta_u(\tau)-\Theta_u(jT)| &\leq (\tau-jT)\kappa~~\forall~u \in \Uc \label{bounding2}
\end{align}
Substituting the above inequalities in~(\ref{alpha-star}), we have
\begin{align}
 &L({\bf G}(\tau+1))-L({\bf G}(\tau))-V \sum_{u \in \Uc}\phi_u(\gamma_u(\tau)) \notag \\
 & \leq \Kc + \left({\bf R}^*
(\tau)-{\boldsymbol \mu}^*(\tau)\right)^\transp\left({\bf Q}(jT)+(\tau-jT)\kappav\right) \notag \\ 
&~~~~~~+\left(\gammav^*(\tau)-\Dm^*(\tau)\right)^\transp\left(\Thetav(jT)+(\tau-jT)\kappav\right) \notag\\
&~~~~~~-V\sum_{u \in \Uc}\phi_u(\gamma_u^*(\tau)). \label{oneslot-arbit}
\end{align}
Then, summing (\ref{oneslot-arbit}) over $\tau \in \{jT, \ldots, (j+1)T-1\}$, we obtain the $T$-slot Lyapunov drift over the $j$-th frame:
\begin{align}
 L&({\bf G}((j+1)T))-L({\bf G}(jT))-V \sum_{\tau=jT}^{jT+T-1}\sum_{u \in \Uc}\phi_u(\gamma_u(\tau)) \notag \\
 & \leq \Kc T+\left(\sum \nolimits_{\tau=jT}^{jT+T-1}\left({\bf R}^*
(\tau)-{\boldsymbol \mu}^*(\tau)\right)\right)^\transp{\bf Q}(jT) \notag \\
&~~~~~~~~+ \left(\sum \nolimits_{\tau=jT}^{jT+T-1}\left({\bf R}^*
(\tau)-{\boldsymbol \mu}^*(\tau)\right)\left(\tau-jT\right)\right)^\transp
\kappav \notag\\
&~~~~~~~~+\left(\sum \nolimits_{\tau=jT}^{jT+T-1}\left(\gammav^*(\tau)-\Dm^*(\tau)\right)\right)^\transp\Thetav(jT) \notag \\
&~~~~~~~~+\left(\sum \nolimits_{\tau=jT}^{jT+T-1}\left(\gammav^*(\tau)-\Dm^*(\tau)\right)\left(\tau-jT\right)\right)^\transp
\kappav \notag \\
&~~~~~~~~-V \sum \nolimits_{\tau=jT}^{jT+T-1}\sum_{u \in \Uc}\phi_u(\gamma_u^*(\tau)) \label{dppopt-comparison}
\end{align}
 Using the inequalities ${\bf R}^*(\tau)-{\boldsymbol \mu}^*(\tau)\leq 2\kappav$,~ 
$\gammav^*(\tau)-\Dm^*(\tau)\leq 2\kappav$ in (\ref{dppopt-comparison}), we have
\begin{align}
 L&({\bf G}((j+1)T))-L({\bf G}(jT))-V \sum_{\tau=jT}^{jT+T-1}\sum_{u \in \Uc}\phi_u(\gamma_u(\tau)) \notag \\
 & \leq \Kc T+\left(\sum \nolimits_{\tau=jT}^{jT+T-1}\left({\bf R}^*
(\tau)-{\boldsymbol \mu}^*(\tau)\right)\right)^\transp{\bf Q}(jT) \notag \\
&~~~~~~~~+ 2\left(\sum \nolimits_{\tau=jT}^{jT+T-1}\left(\tau-jT\right)\right)\kappav^\transp\kappav \notag\\
&~~~~~~~~+\left(\sum \nolimits_{\tau=jT}^{jT+T-1}\left(\gammav^*(\tau)-\Dm^*(\tau)\right)\right)^\transp\Thetav(jT)\notag \\
&~~~~~~~~+2\left(\sum \nolimits_{\tau=jT}^{jT+T-1}\left(\tau-jT\right)\right)\kappav^\transp\kappav \notag \\
&~~~~~~~~-V\sum \nolimits_{\tau=jT}^{jT+T-1} \sum_{u \in \Uc}\phi_u(\gamma_u^*(\tau))
\end{align} 
Using $\kappav^\transp\kappav \leq \frac{\Kc}{2}$, $\sum_{\tau=jT}^{jT+T-1}(\tau-jT)=\frac{T(T-1)}{2}$, we get
\begin{align}
 &L({\bf G}((j+1)T))-L({\bf G}(jT))-V\sum_{\tau=jT}^{jT+T-1}\sum_{u \in \Uc}\phi_u(\gamma_u(\tau)) \leq \notag \\
 & \Kc T+\Kc T(T-1) +\left(\sum \nolimits_{\tau=jT}^{jT+T-1}\left({\bf R}^*
(\tau)-{\boldsymbol \mu}^*(\tau)\right)\right)^\transp{\bf Q}(jT) \notag \\
&~+\left(\sum \nolimits_{\tau=jT}^{jT+T-1}\left(\gammav^*(\tau)-\Dm^*(\tau)\right)\right)^\transp\Thetav(jT) \notag \\
&~-V\sum \nolimits_{\tau=jT}^{jT+T-1} \sum_{u \in \Uc}\phi_u(\gamma_u^*(\tau))\label{common-allpolicies}
\end{align}
We  now consider the policy $\{a^*(\tau)\}_{\tau=jT}^{(j+1)T-1}$  satisfying the following constraints:\footnote{It is easy to see that such policy is guaranteed to exist 
provided that we allow, without loss of generality, for a virtual video layer of zero quality and zero rate, and in the assumption that, at any time $t$, 
each user $u$ has at least one link $(h,u) \in \Ec$ with $h \in \Nc(u) \cap \Hc(f_u)$ with peak rate $C_{hu}(t)$ lower bounded by some strictly positive number $C_{\min}$.
This prevents the case where a user gets zero rate for a whole frame of length $T$. This assumption is not restrictive in practice since a user experiencing unacceptably 
poor link quality to all the helpers for a long time interval would be disconnected from the network and its streaming session halted.}
\begin{align}
&\frac{1}{T} \sum_{\tau=jT}^{(j+1)T-1}\left[
kR^*_{hu}\left(\tau\right)-n\mu^*_{hu}\left(\tau\right)\right] < -\epsilon~\forall~(h, u) \in \Ec \label{primeineq-1}  \\
& \frac{1}{T} \sum_{\tau=jT}^{(j+1)T-1}\left[
\gamma^*_u\left(\tau\right)-D_u^*\left(\tau\right)\right] < -\epsilon~\forall~ u \in \Uc \label{primeineq-2}
\end{align}
where $\epsilon>0$ is arbitrary. 
We plug in the inequalities (\ref{primeineq-1}), (\ref{primeineq-2}) in 
(\ref{common-allpolicies}) and obtain
\begin{align}
 L&({\bf G}((j+1)T))-L({\bf G}(jT))-V\sum_{\tau=jT}^{jT+T-1}\sum_{u \in \Uc}\phi_u(\gamma_u(\tau))< \notag \\
 &  \Kc T^2 -\epsilon T \sum_{(h,u) \in \Ec}Q_{hu}(jT) - \epsilon T \sum_{u \in \Uc}\Theta_{u}(jT)\notag \\
 &-V\sum \nolimits_{\tau=jT}^{jT+T-1} \sum_{u \in \Uc}\phi_u(\gamma_u^*(\tau))
\end{align}
Also, considering the fact that for any vector $\gammav = (\gamma_1, \ldots, \gamma_{|\Uc|})$ we have
\begin{align}
\sum_{u \in \Uc}\phi_u(D_u^{\min}) = \phi_{\min} \leq \sum_{u \in \Uc}\phi_u(\gamma_u) &\leq \phi_{\max} \notag \\
&= \sum_{u \in \Uc}\phi_u(D_u^{\max}),
\end{align} 
we can write:
\begin{align}
 &L({\bf G}((j+1)T))-L({\bf G}(jT)) < \notag \\
 &\Kc T^2 +VT(\phi_{\max}-\phi_{\min}) - \epsilon T \sum_{(h,u) \in \Ec}Q_{hu}(jT) \notag \\
 &~~~~~~~~~~~~~~~~~~~~~~~~~~~~~~~-\epsilon T \sum_{u \in \Uc}\Theta_{u}(jT) 
\end{align}
Once again using (\ref{bounding1}), (\ref{bounding2}), we have:
\begin{align}
& L({\bf G}((j+1)T))-L({\bf G}(jT)) <\notag \\
& \Kc T^2 +VT(\phi_{\max}-\phi_{\min}) - \epsilon \sum_{\tau=jT}^{jT+T-1} \sum_{(h,u) \in \Ec}Q_{hu}(\tau) \notag \\
&~~~~~-\epsilon \sum_{\tau=jT}^{jT+T-1} \sum_{u \in \Uc}\Theta_{u}(\tau) + \frac{\epsilon  \kappa (|\Ec|+|\Uc|)T(T-1)}{2}
\end{align}
Summing the above over the frames $j \in \{0, \ldots, F-1\}$ yields
\begin{align}
&L({\bf G}((FT))-L({\bf G}(0)) < \notag \\
& \Kc T^2F + VFT(\phi_{\max}-\phi_{\min}) - \epsilon \sum_{\tau=0}^{FT-1} \sum_{(h,u) \in \Ec}Q_{hu}(\tau) \notag \\
&~~~~~~~ - \epsilon \sum_{\tau=0}^{FT-1} \sum_{u \in \Uc}\Theta_{u}(\tau) + \frac{\epsilon  \kappa (|\Ec|+|\Uc|)FT(T-1)}{2}
\end{align}
Rearranging and neglecting appropriate terms, we get
\begin{align}
  &\frac{1}{FT} \sum_{\tau=0}^{FT-1} \sum_{(h,u) \in \Ec}Q_{hu}(\tau) 
+ \frac{1}{FT}\sum_{\tau=0}^{FT-1} \sum_{u \in \Uc}\Theta_{u}(\tau) <  \notag \\
& \frac{\Kc T}{\epsilon} + \frac{V(\phi_{\max}-\phi_{\min})}{\epsilon}+ \frac{L({\bf G}(0))}{\epsilon FT} + \frac{ \kappa (|\Ec|+|\Uc|)(T-1)}{2}
\end{align}
Taking limits as $F \rightarrow \infty$
\begin{empheq}[box = \fbox]{align}
 & \lim_{F \rightarrow \infty} \frac{1}{FT} \sum_{\tau=0}^{FT-1}\left( \sum_{(h,u) \in \Ec}Q_{hu}(\tau) 
+ \sum_{u \in \Uc}\Theta_{u}(\tau)\right) < \notag \\
 & \frac{\Kc T}{\epsilon} + \frac{V(\phi_{\max}-\phi_{\min})}{\epsilon}+ \frac{ \kappa (|\Ec|+|\Uc|)(T-1)}{2} \label{strongstab-arbit}
\end{empheq}
such that (\ref{strongstab-arbit1}) is proved.  

We now consider the policy $\{a^*(\tau)\}_{\tau=jT}^{(j+1)T-1}$ which achieves the optimal solution $\phi_j^{\rm opt}$ to the problem (\ref{constraint}) -- (\ref{feasactions-trans-arbit}). Using (\ref{qstable-arbit}) and (\ref{virt-stable-arbit}) in (\ref{common-allpolicies}), we have 
\begin{align}
 &L({\bf G}((j+1)T))-L({\bf G}(jT))-V\sum_{\tau=jT}^{jT+T-1}\sum_{u \in \Uc}\phi_u(\gamma_u(\tau))\leq\notag \\
 &~~~~~~ \Kc T+\Kc T(T-1)-V T\phi_j^{\rm opt}
\end{align}
Summing this over $j \in \{0, \ldots, F-1\}$, yields
\begin{align}
L({\bf G}&((FT))-L({\bf G}(0))-V \sum_{\tau=0}^{FT-1}\sum_{u \in \Uc}\phi_u(\gamma_u(\tau))\leq \notag \\
&\Kc T^2F-V T\sum_{j=0}^{F-1}\phi_j^{\rm opt}.
\end{align}
Dividing both sides by $VFT$ and using the fact that $L({\bf G}((FT)) > 0$ , we get 
\begin{align}
\frac{1}{FT}\sum_{\tau=0}^{FT-1}\sum_{u \in \Uc}\phi_u(\gamma_u(\tau)) \geq \frac{1}{F}\sum_{j=0}^{F-1}\phi_j^{\rm opt} - \frac{\Kc T}{V} - \frac{L({\bf G}(0))}{VTF}.
\end{align}
At this point, using Jensen's inequality, the fact that $\phi_u(\cdot)$ is continuous and non-decreasing for all $u \in \Uc$, 
and the fact that the strong stability of the queues (\ref{strongstab-arbit}) implies that 
$\lim_{F \rightarrow \infty} \frac{1}{FT} \sum_{\tau=0}^{FT-1}\Theta_{u}(\tau) < \infty ~\forall~ u \in \Uc$,  
which in turns implies that $\overline{\gamma}_u \leq \overline{D}_u~\forall~u \in \Uc$, we arrive at 
\begin{align}
\boxed{\sum_{u \in \Uc}\phi_u\left(\overline{D}_u\right) \geq
 \lim_{F \rightarrow \infty}\frac{1}{F}\sum_{j=0}^{F-1}\phi_j^{\rm opt} - \frac{\Kc T}{V}.}\label{optutil-arbit}
\end{align}
such that (\ref{optutil-arbit1}) is proved.  

Thus, the  utility under the DPP policy is within $O(1/V)$ of the time average of the $\phi_j^{\rm opt}$ utility values that 
can be achieved only  if knowledge of the future states up to a look-ahead of blocks of $T$ slots.  
If $T$ is increased, then the value of $\phi_j^{\rm opt}$ for every frame $j$ improves since we allow a larger 
look-ahead.  However, from (\ref{optutil-arbit}), we can see that if $T$ is increased, then $V$ can also be increased 
in order to maintain the same distance from optimality. This yields a corresponding $O(V)$ increase in the queues backlog. 

For the case where the network state $\omegav(t)$ is stationary and ergodic, the time average in the left hand side of (\ref{strongstab-arbit}) 
and in the right hand side of (\ref{optutil-arbit})  become ensemble averages because of ergodicity. Thus, we obtain 
(\ref{utilperf-iid}) and (\ref{strongstab-iid}). Furthermore, if the network state is i.i.d., we can take $T = 1$ in the above derivation, obtaining 
the bounds given in Corollary  \ref{corcor}. 
\bibliographystyle{IEEEtran}
\bibliography{betharef}

\end{document}

%% file: bethamacros.tex
\long\def\comment#1{}

\newfont{\bbb}{msbm10 scaled 700}

\newfont{\bb}{msbm10 scaled 1100}

\newcommand{\Dm}{{\bf D}}

\newcommand{\Gm}{{\bf G}}

\newcommand{\Qm}{{\bf Q}}
\newcommand{\Rm}{{\bf R}}

\newcommand{\Ac}{{\cal A}}

\newcommand{\Cc}{{\cal C}}

\newcommand{\Ec}{{\cal E}}
\newcommand{\Fc}{{\cal F}}
\newcommand{\Gc}{{\cal G}}
\newcommand{\Hc}{{\cal H}}

\newcommand{\Kc}{{\cal K}}

\newcommand{\Nc}{{\cal N}}

\newcommand{\Rc}{{\cal R}}

\newcommand{\Uc}{{\cal U}}

\newcommand{\gammav}{\hbox{\boldmath$\gamma$}}

\newcommand{\muv}{\hbox{\boldmath$\mu$}}

\newcommand{\Thetav}{\hbox{\boldmath$\Theta$}}

\newcommand{\omegav}{\hbox{\boldmath$\omega$}}

\newcommand{\kappav}{\hbox{\boldmath$\kappa$}}

\newcommand{\Thetam}{\hbox{\boldmath$\Theta$}}

\newcommand{\transp}{{\sf T}}

